\documentclass[journal,twoside,romanappendices]{IEEEtran}
\pdfoutput=1

\usepackage{amsmath}
\usepackage{amssymb}
\usepackage{amsfonts}
\usepackage{amscd}
\usepackage{amsthm}
\usepackage[noadjust]{cite}
\usepackage{threeparttable}
\usepackage{textcomp}
\usepackage{multirow}

\usepackage{mathrsfs}

\usepackage{paralist}

\newcommand{\sfrac}{\frac}

\usepackage[caption=false,font=footnotesize]{subfig}

\newtheorem{theorem}{Theorem}
\newtheorem{lemma}{Lemma}

\theoremstyle{definition}
\newtheorem{example}{Example}
\newtheorem{definition}{Definition}

\makeatletter
\renewcommand*\env@matrix[1][c]{\hskip -\arraycolsep
  \let\@ifnextchar\new@ifnextchar
  \array{*\c@MaxMatrixCols #1}}
\makeatother
\usepackage{graphicx}

\newcommand{\La}{\Lambda}
\newcommand{\Lt}{\widetilde{\Lambda}}
\newcommand{\Rb}{\mathbb{R}}
\newcommand{\Zb}{\mathbb{Z}}
\newcommand{\Cb}{\mathbb{C}}
\newcommand{\Quat}{\mathcal{Q}}
\newcommand{\Hb}{\mathbb{H}}
\newcommand{\Db}{\mathbb{D}}
\newcommand{\Ls}{\Lambda_{\rm c}}
\newcommand{\Lts}{\widetilde{\Lambda}_{\rm c}}
\newcommand{\Is}{I_{\rm c}}

\newcommand{\real}{{\rm Re}}
\newcommand{\imag}{{\rm Im}}
\newcommand{\vol}{{\rm Vol}}
\newcommand{\vect}{{\rm vec}}
\newcommand{\divides}{\,\vert\,}
\newcommand{\om}{\omega}
\newcommand{\omb}{\overline{\omega}}
\newcommand{\nbytwo}{\sfrac{n}{2}}

\newcommand{\Dc}{\mathcal{D}}
\newcommand{\Cc}{\mathscr{C}}
\newcommand{\modc}{{\rm~mod~}}
\newcommand{\Lscomb}{\Lambda_{\rm c}^{(12)}}
\newcommand{\Lacomb}{\Lambda_{12}}
\newcommand{\Lsth}{\Lambda_{\rm c}^{(3)}}
\newcommand{\Lath}{\Lambda_3}

\title{Lattice Index Coding}
\author{
\IEEEauthorblockN{Lakshmi~Natarajan, Yi~Hong,~\IEEEmembership{Senior~Member,~IEEE}, and Emanuele~Viterbo,~\IEEEmembership{Fellow,~IEEE}}%
\thanks{This work was supported by the Australian Research Council (ARC) Discovery Project (ARC~DP130100103). The material in this paper was presented in part at the IEEE Information Theory Workshop, Jerusalem, April~26--May~01, 2015.}%
\thanks{The authors are with the Department of Electrical and Computer System Engineering, Monash University, VIC 3800, Australia (e-mail: \{lakshmi.natarajan, yi.hong, emanuele.viterbo\}@monash.edu).}%
\thanks{Copyright~\copyright~2014~IEEE. Personal use of this material is permitted. However, permission to use this material for any other purposes must be obtained from the IEEE by sending a request to pubs-permissions@ieee.org.}
}

\begin{document}

\maketitle

\begin{abstract}
\boldmath
The index coding problem involves a sender with $K$ messages to be transmitted across a broadcast channel, and a set of receivers each of which demands a subset of the $K$ messages while having prior knowledge of a different subset as side information. We consider the specific case of noisy index coding where the broadcast channel is Gaussian and every receiver demands all the messages from the source. Instances of this communication problem arise in wireless relay networks, sensor networks, and retransmissions in broadcast channels. We construct \emph{lattice index codes} for this channel by encoding the $K$ messages individually using $K$ modulo lattice constellations and transmitting their sum modulo a coarse lattice. We introduce a design metric called \emph{side information gain} that measures the advantage of a code in utilizing the side information at the receivers, and hence, its goodness as an index code. Based on the Chinese remainder theorem, we then construct lattice index codes with large side information gains using lattices over the following principal ideal domains: rational integers, Gaussian integers, Eisenstein integers, and the Hurwitz quaternions. Among all lattice index codes constructed using any densest lattice of a given dimension, our codes achieve the maximum side information gain. Finally, using an example, we illustrate how the proposed lattice index codes can benefit Gaussian broadcast channels with more general message demands.
\end{abstract}


\section{Introduction} \label{sec:1}

\IEEEPARstart{T}{he} classical noiseless index coding problem consists of a sender with $K$ independent messages \mbox{$w_1,\dots,w_K$}, and a noiseless broadcast channel, where each receiver demands a subset of the messages, while knowing the values of a different subset of messages as side information. The transmitter is required to broadcast a coded packet, with the least possible length, to meet the demands of all the receivers~(see~\cite{BiK_INFOCOM_98,YBJK_IEEE_IT_11,ALSWH_FOCS_08,RSG_IEEE_IT_10,BKL_arxiv_10,UnW_ISIT_13} and references therein).
In the noisy version of this problem, the messages are to be transmitted across a broadcast channel with additive white Gaussian noise (AWGN) at the receivers~(see \cite{Wu_ISIT_07,KrS_ITW_07,Xie_CWIT_07,OSBB_IEEE_IT_08,OWT_ISIT_12,YLX_ISIT_09,SiC_ISIT_14,AOJ_ISIT_14,Tun_IEEE_IT_06} and references therein). The exact capacity region (the achievable rates of the $K$ messages) with general message demands and side informations is known only for the two-receiver case~\cite{Wu_ISIT_07,KrS_ITW_07}. 

\begin{figure*}[t!]
\centering
\subfloat[Phase 1: ${\rm BS}_1$ and ${\rm BS}_2$ transmit files $w_1$ and $w_2$.]{\includegraphics[width=3.5in]{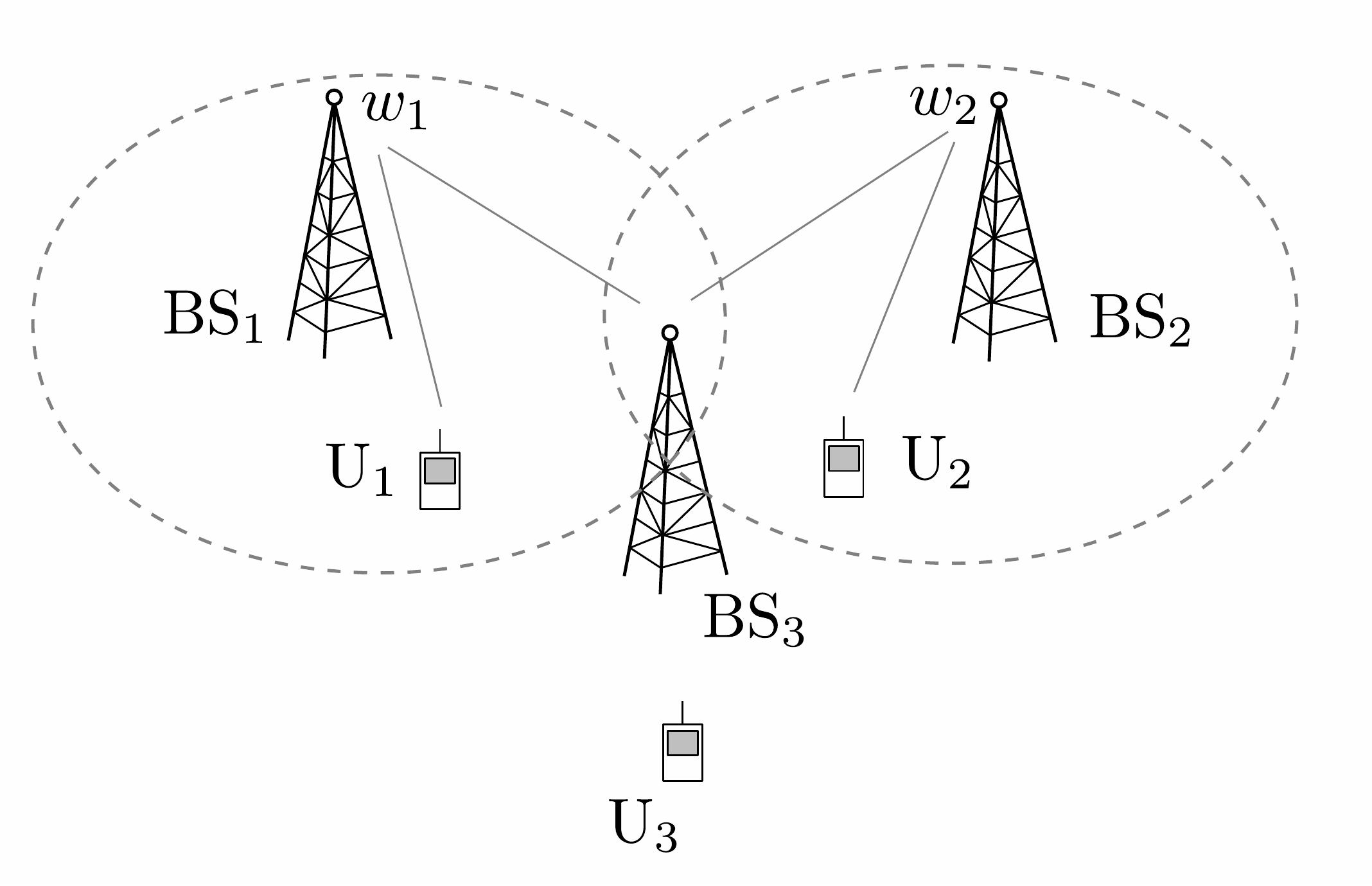}}
\hfil
\subfloat[Phase 2: ${\rm BS}_3$ transmits $w_1$ and $w_2$ using noisy index coding.]{\includegraphics[width=3.5in]{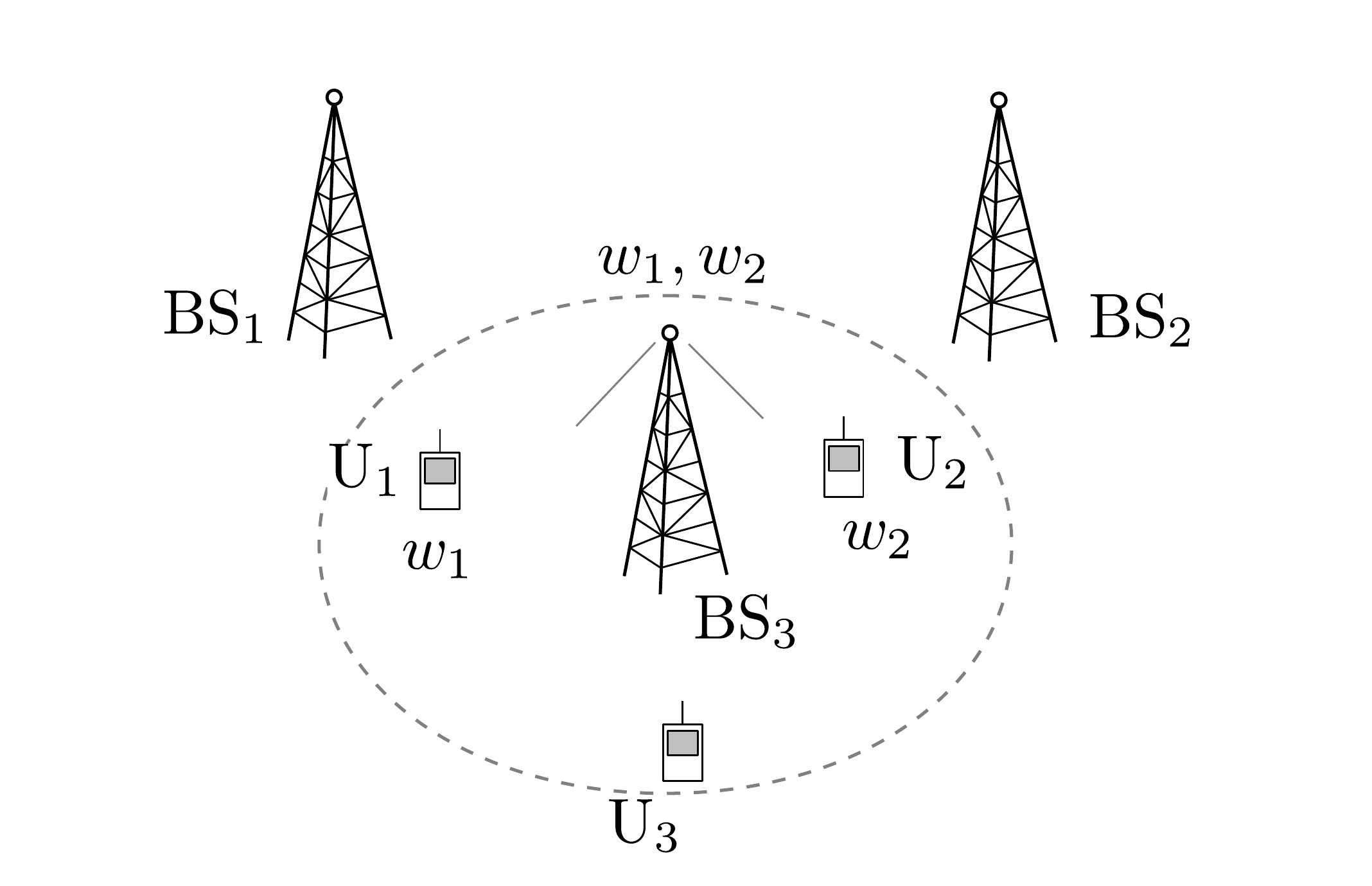}}
\vspace{2mm}
\caption{Common message broadcast with receiver side information in the wireless `butterfly' network: (a) ${\rm BS}_1$ and ${\rm BS}_2$ simultaneously broadcast files $w_1$ and $w_2$. At the end of Phase~1, ${\rm U}_1$ receives $w_1$, ${\rm U}_2$ receives $w_2$, and ${\rm BS}_3$ receives both. (b) In Phase~2, ${\rm BS}_3$ transmits $w_1,w_2$ using noisy index coding to utilize side information at ${\rm U}_1$ and ${\rm U}_2$ while being intelligible to ${\rm U}_3$.}
\label{fig:cellular}
\end{figure*}

We consider the special case of noisy index coding where every receiver demands all the messages at the source.
Instances of this communication problem are encountered in wireless relay networks~\cite{OSBB_IEEE_IT_08,KrS_ITW_07,Xie_CWIT_07}, retransmissions in broadcast channels~\cite{BiK_INFOCOM_98}, and communications in sensor networks~\cite{Tun_IEEE_IT_06}.
Fig.~\ref{fig:cellular} illustrates a wireless version of the `butterfly' network where noisy index coding is useful.
Two data packets $w_1$ and $w_2$, which are available at the base stations ${\rm BS}_1$ and ${\rm BS}_2$, respectively, are to be broadcast to all three users ${\rm U}_1,{\rm U}_2,{\rm U}_3$ in the network through a decode-and-forward helper node ${\rm BS}_3$. The nodes ${\rm U}_1$ and ${\rm BS}_3$ are within the range of ${\rm BS}_1$, ${\rm U}_2$ and ${\rm BS}_3$ are within the range of ${\rm BS}_2$, and all three users are in the range of ${\rm BS}_3$. In the first phase of the protocol, both ${\rm BS}_1$ and ${\rm BS}_2$ simultaneously broadcast their corresponding data packets. While ${\rm U}_1$ and ${\rm U}_2$ decode $w_1$ and $w_2$, respectively, the helper node ${\rm BS}_3$ experiences a multiple-access channel and decodes both the messages. In the second phase of the protocol, ${\rm BS}_3$ broadcasts $w_1$ and $w_2$ to all three users. While ${\rm U}_1$ and ${\rm U}_2$ are aided by the data packets received in the first phase of the protocol, no such side information is available at ${\rm U}_3$.
The traditional approach of broadcasting the bit-wise XOR of $w_1$ and $w_2$ in the second phase is not useful, since it does not satisfy the demands of ${\rm U}_3$. On the other hand, performing index coding at the physical layer will allow us to convert the side informations at ${\rm U}_1$ and ${\rm U}_2$ into performance gains while meeting the demands of all three receivers.

\begin{figure*}[t!]
\centering
\subfloat[Original broadcast phase.]{\includegraphics[width=3.5in]{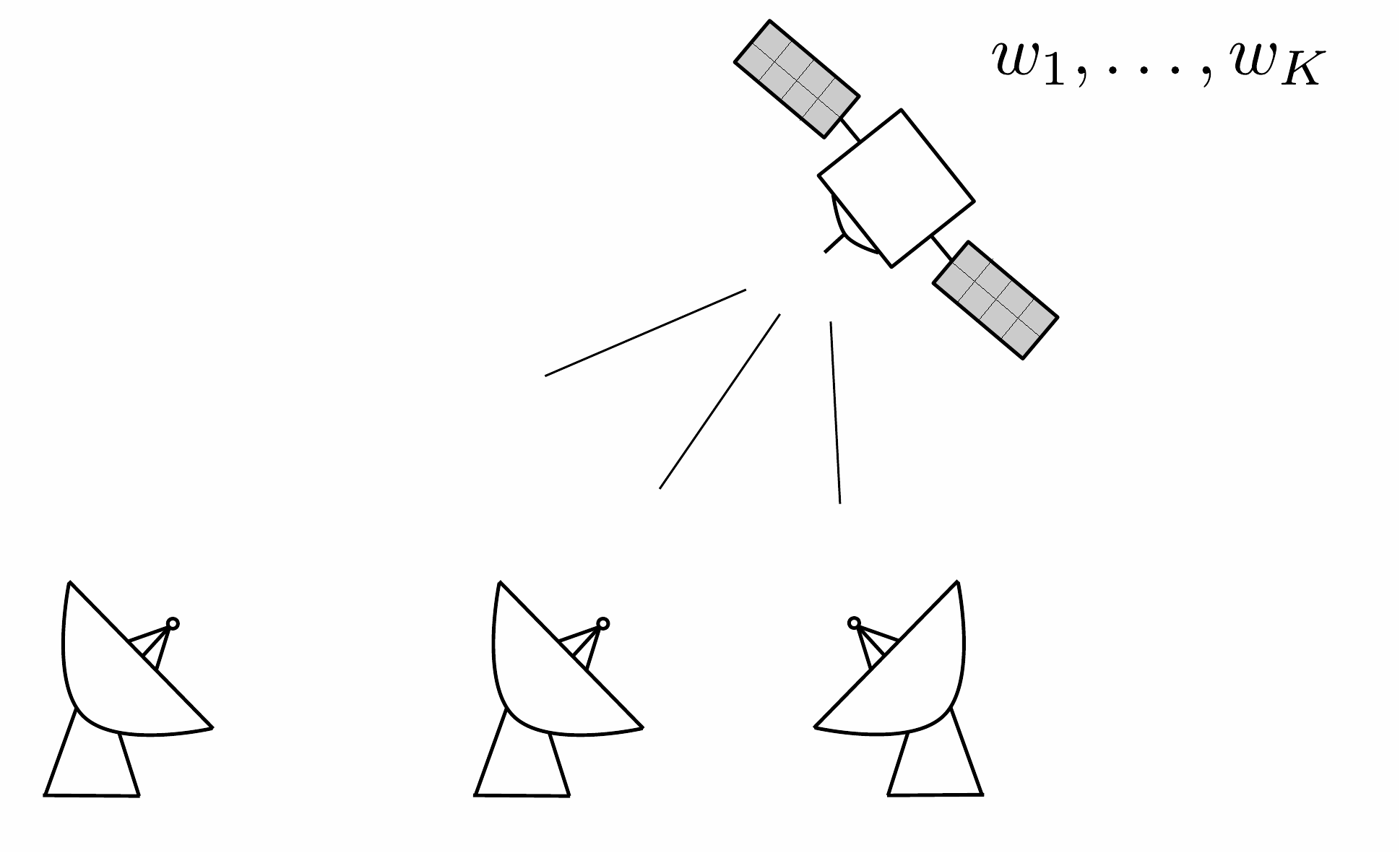}}
\hfil
\subfloat[Retransmission phase: receivers have side information.]{\includegraphics[width=3.5in]{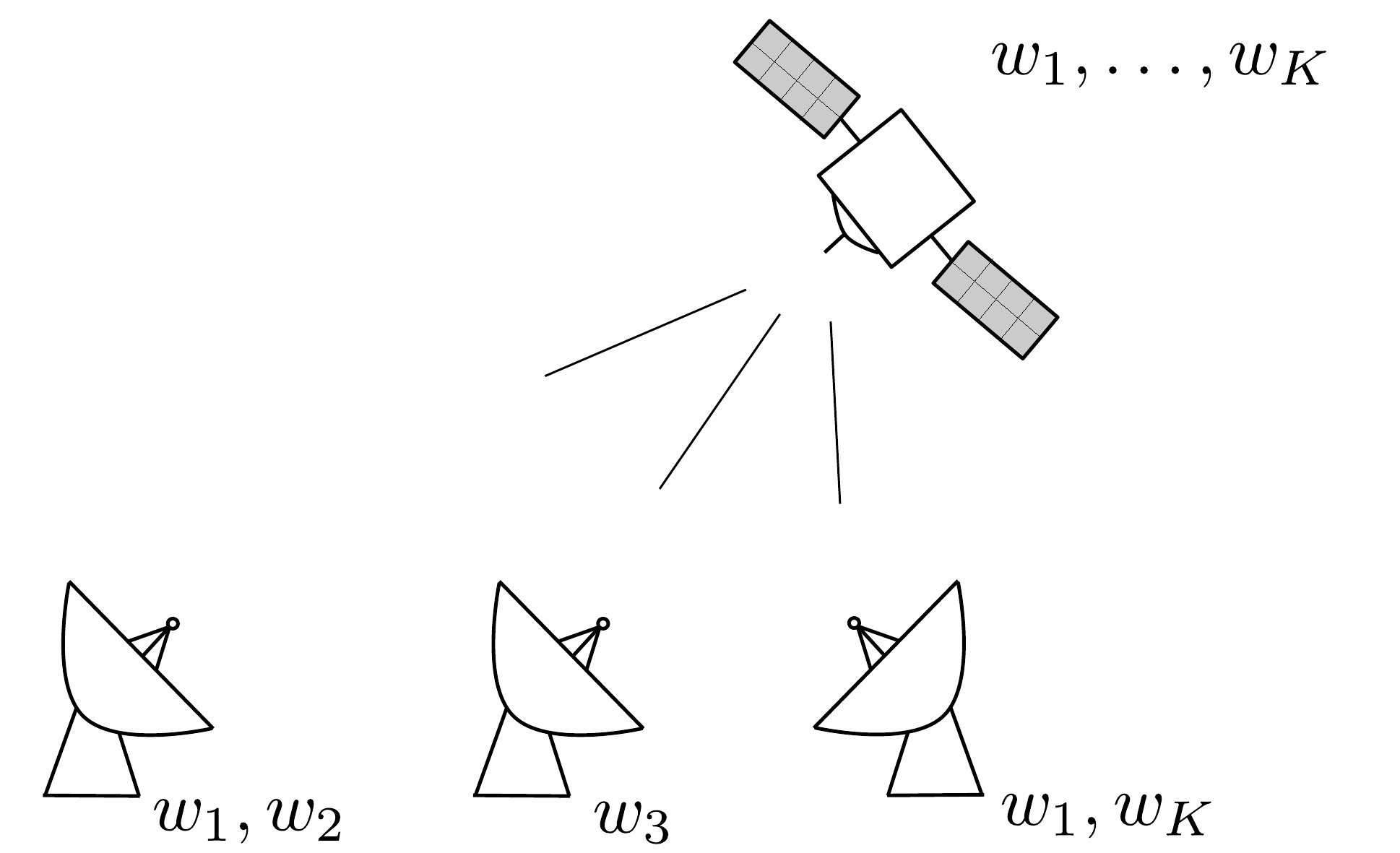}}
\vspace{2mm}
\caption{Common message broadcast with receiver side information in satellite communications: (a) The satellite broadcasts a common message containing $K$ data packets to multiple terrestrial receivers. Due to intermittent channel variations, each receiver successfully decodes only a subset of the $K$ packets. Here, the first receiver decodes $w_1,w_2$, the second $w_3$, and the third $w_1,w_K$. (b) In the retransmission phase the satellite performs noisy index coding to exploit this side information at the receivers.}
\label{fig:satellite}
\end{figure*}

Noisy index coding for broadcasting common messages is also useful in the retransmission phase of satellite broadcasting services, which was the original motivation for considering (noiseless) index codes~\cite{BiK_INFOCOM_98}. Consider a satellite downlink, as shown in Fig.~\ref{fig:satellite}, where a common message consisting of $K$ data packets is broadcast to multiple terrestrial receivers. Due to varying channel conditions, each receiver successfully decodes (possibly different) parts of the transmitted frame. In the retransmission phase of the protocol, the satellite can use a noisy index code to simultaneously broadcast the $K$ packets while exploiting the side informations at all the receivers.

\subsection{Background}

The capacity region of the common message Gaussian broadcast channel with receiver side information follows from the results in~\cite{Tun_IEEE_IT_06}. Denote a receiver by $({\sf SNR},S)$, where ${\sf SNR}$ is the signal-to-noise ratio, and \mbox{$S\subset\{1,\dots,K\}$} is the index set of the messages \mbox{$w_S=(w_k,k \in S)$} whose values are known at the receiver as side information.
Note that this terminology includes the case \mbox{$S=\varnothing$}, i.e., no side information.
Let \mbox{$R_1,\dots,R_K$} be the rates of the individual messages in bits per dimension (b/dim), i.e., the number of bits to be transmitted per each use of the broadcast channel.
The source entropy is \mbox{$R=R_1+\cdots+R_K$}, and the \emph{side information rate} at \mbox{$({\sf SNR},S)$} is defined as \mbox{$R_S \triangleq \sum_{k \in S}R_k$}.
The rate tuple $(R_1,\dots,R_K)$ is achievable if and only if~\cite{Tun_IEEE_IT_06}
\begin{align*} 
\frac{1}{2}\log_2 \left( 1 + {\sf SNR} \right) > H(w_1,\dots,w_K \, \vert \, w_S) = R - R_S,
\end{align*}
for every receiver $({\sf SNR},S)$.
Consequently, at high message rates, the presence of the side information corresponding to $S$ at a receiver reduces the minimum required ${\sf SNR}$ from approximately $2^{2R}$ to $2^{2(R-R_S)}$, or equivalently, by a factor of \mbox{$R_S \times 20\log_{10} 2 \text{ dB} \approx 6 R_S \text{ dB}$}. Hence, a capacity-achieving index code allows a receiver to transform each bit per dimension of side information into an apparent ${\sf SNR}$ gain of approximately \mbox{$6$~dB}.

The notion of \emph{multiple interpretation} was introduced in~\cite{XFKC_CISS_06} as a property of error correcting codes that allows the receiver performance to improve with the availability of side information.
Binary multiple interpretation codes based on nested convolutional and cyclic codes were constructed in~\cite{MLV_PIMRC_12} and~\cite{BaC_ITW_11}, respectively.
These codes can be viewed as index codes for the noisy binary broadcast channel.
To the best of our knowledge, there has been no prior work in designing index codes for the AWGN broadcast channel.

\subsection{Contributions}

In this work, we propose \emph{lattice index codes} $\mathscr{C}$ for the AWGN broadcast channel, in which the $K$ messages are individually mapped to $K$ modulo lattice constellations, and the transmit symbol is generated as the sum of the individual symbols modulo a coarse lattice.

Given the value of $w_S$ as side information, the optimal decoder restricts its choice of symbols to a subset of $\mathscr{C}$, thereby increasing the minimum squared Euclidean distance between the valid codewords. We use this squared distance gain, normalized by the side information rate $R_S$, as the design metric, and call it the \emph{side information gain} of the code $\mathscr{C}$.
We first motivate our results using a simple one-dimensional lattice code over $\Zb$ (Section~\ref{sec:motivating_example}), and then show that \mbox{$20 \log_{10} 2 \approx 6$~dB/b/dim} is an upper bound on the side information gain of lattice index codes constructed from densest lattices~(Section~\ref{sec:linear_lattice_codes}).
Note that this upper bound characterizes the maximum squared distance gain, and is independent of the information theoretic result of~\cite{Tun_IEEE_IT_06} which characterizes the ${\sf SNR}$ gain asymptotically in both the code dimension and probability of error.
Based on the Chinese remainder theorem, we construct index codes for the AWGN channel using lattices over the following principal ideal domains~(PIDs): rational integers $\Zb$, Gaussian integers $\Zb[i]$, Eisenstein integers $\Zb[\omega]$, and the Hurwitz quaternion integers $\Hb$~(Sections~\ref{sec:commutative} and~\ref{sec:Hurwitz}).
All the proposed lattice index codes provide a side information gain of \mbox{$20 \log_{10} 2$~dB/b/dim}.
Among all lattice index codes constructed using the densest lattices in any given dimension, our codes provide the optimal side information gain.
Finally, using the example of a three receiver Gaussian broadcast channel with private message requests, we illustrate how the proposed lattice index codes can be utilized under more general message demands (Section~\ref{sec:3receiver}).


\subsection{Recent results}

Since the submission of the initial version of this paper, further results on index codes for the common message Gaussian broadcast channel have been reported.
The lattice index codes presented in this paper are designed using tuples of distinct prime numbers, and hence, the resulting rates of the $K$ messages are not all equal to each other, and the alphabet sizes of the messages are not powers of $2$. New lattice index codes are reported in~\cite{Hua_ISIT_15} that generalize the $\Zb[i]$ and $\Zb[\omega]$ based constructions of Section~\ref{sec:commutative} to arbitrary algebraic number fields. Further,~\cite{Hua_ISIT_15} constructs sequences of lattice index codes, that consist of one code for each value of $K$, for encoding all the $K$ messages at the same rate.
Index codes based on multidimensional pulse amplitude modulation (PAM) constellations have been obtained in~\cite{NHV_Comml_15}
that encode all the messages at the same rate and allow alphabet sizes that are powers of $2$. In~\cite{NHV_ISIT_15}, the achievable rate region of a concatenated coding scheme that uses an inner index code for modulation and $K$ independent outer channel codes for noise resilience has been analyzed. This concatenated scheme has been shown to convert the noisy index coding channel into a multiple-access channel and perform close to the channel capacity.


\emph{Notation:} 
We use \mbox{$i=\sqrt{-1}$} and \mbox{$\omega=\exp\left( \sfrac{i 2\pi}{3} \right)$}. The symbol $S^\mathsf{c}$ denotes the complement of the set $S$, and $\varnothing$ is the empty set. For a complex number $m$, the symbols $\overline{m}$, $\real(m)$ and $\imag(m)$ denote the conjugate, the real part, and the imaginary part of $m$, respectively. The operator $(\cdot)^\intercal$ is the transpose of a matrix or a vector, and $\|\cdot\|$ is the Euclidean norm of a vector.

\section{Motivating Example} \label{sec:motivating_example}

The lattice index codes proposed in Sections~\ref{sec:commutative} and~\ref{sec:Hurwitz} achieve a large side information gain by providing a squared distance gain that is exponential in the side information rate $R_S$ for \mbox{$S \subset \{1,\dots,K\}$}. In this section, we illustrate the key idea behind our construction using a simple one-dimensional lattice index code~(Example~\ref{ex:toy_example}).

Let \mbox{$w_1,\dots,w_K$} be $K$ independent messages at the source with alphabets \mbox{$\mathcal{W}_1,\dots,\mathcal{W}_K$}, respectively.
The transmitter jointly encodes the information symbols \mbox{$w_1,\dots,w_K$}, to a codeword \mbox{$x \in \mathscr{C}$}, where \mbox{$\mathscr{C} \subset \Rb^n$} is an $n$-dimensional constellation.
The rate of the $k^{\text{th}}$ message is \mbox{$R_k = \sfrac{1}{n} \log_2 |\mathcal{W}_k|$}~b/dim, \mbox{$k=1,\dots,K$}.
Given the channel output \mbox{$y=x+z$}, where $z$ is the additive white Gaussian noise, and the side information $\mbox{$w_S=a_S$}$, i.e., $w_k=a_k$ for $k \in S$, the maximum-likelihood decoder at the receiver $({\sf SNR},S)$ restricts its search to the subcode \mbox{$\mathscr{C}_{a_S} \subset \mathscr{C}$} obtained by expurgating all the codewords in $\mathscr{C}$ that correspond to \mbox{$w_S \neq a_S$}.
Denote the minimum distance between any two points in $\mathscr{C}$ by $d_0$. Let $d_{a_S}$ be the minimum distance of the subcode $\mathscr{C}_{a_S}$, and $d_S$ be the minimum of $d_{a_S}$ over all possible values \mbox{$a_S$} of side information $w_S$. Then the minimum squared distance gain corresponding to the side information index set $S$ is $10 \log_{10} \left( \sfrac{d_S^2}{d_0^2} \right)$~dB. 

The performance improvement at the receiver due to $S$ is observed as a shift in the probability of error curve (versus ${\sf SNR}$) to the left.
The squared distance gain $10 \log_{10} \left(\sfrac{d_S^2}{d_0^2}\right)$~dB is a first-order estimate of this apparent ${\sf SNR}$ gain.
Normalizing with respect to the side information rate \mbox{$R_S = \sum_{k \in S}R_k$}, and minimizing over all subsets $S$, we see that each bit per dimension of side information provides a squared distance gain of at least
\begin{align} \label{eq:side_inf_gain}
\Gamma(\mathscr{C}) \triangleq \min_{S} \frac{10 \log_{10}\left( \sfrac{d_S^2}{d_0^2} \right)}{R_S}.
\end{align}
We call $\Gamma(\mathscr{C})$ the \emph{side information gain} of the code $\mathscr{C}$, and its unit is $\text{dB/b/dim}$.

For a given code $\mathscr{C}$, the gain available from $S$ is at least $R_S \times \Gamma(\mathscr{C})$~dB with respect to the baseline performance of $\mathscr{C}$ in the classical point-to-point AWGN channel, i.e., with no side information.
For $\mathscr{C}$ to be a good index code for the AWGN broadcast channel, we require that
\begin{inparaenum}[1)]
\item $\mathscr{C}$ be a good point-to-point AWGN code, in order to minimize the ${\sf SNR}$ requirement at the receiver with no side information; and
\item $\Gamma(\mathscr{C})$ be large, so as to maximize the minimum gain from the availability of side information at the other receivers.
\end{inparaenum}

An additional desirable property is that the normalized gain ${10 \log_{10}\left(\sfrac{d_S^2}{d_0^2}\right)}/{R_S}$ provided by the lattice index code be constant for every $S$, i.e.,
\begin{align} \label{eq:uniform_gain}
\Gamma\left( \mathscr{C} \right) = \frac{10 \log_{10}\left(\sfrac{d_S^2}{d_0^2}\right)}{R_S} \, \text{ for every } S \subset \{1,\dots,K\}.
\end{align}
We say that a lattice index code provides \emph{uniform gain} if it satisfies~\eqref{eq:uniform_gain}.
A necessary and sufficient condition for a lattice index code to be a uniform gain code is that $d_S$ is exponential in $R_S$.
All the index codes constructed in Sections~\ref{sec:commutative} and~\ref{sec:Hurwitz} are uniform gain lattice index codes with $\Gamma(\mathscr{C}) \approx 6$~dB/b/dim.

\begin{example} \label{ex:toy_example}
\begin{figure*}[!t]
\centering
\includegraphics[width=5in]{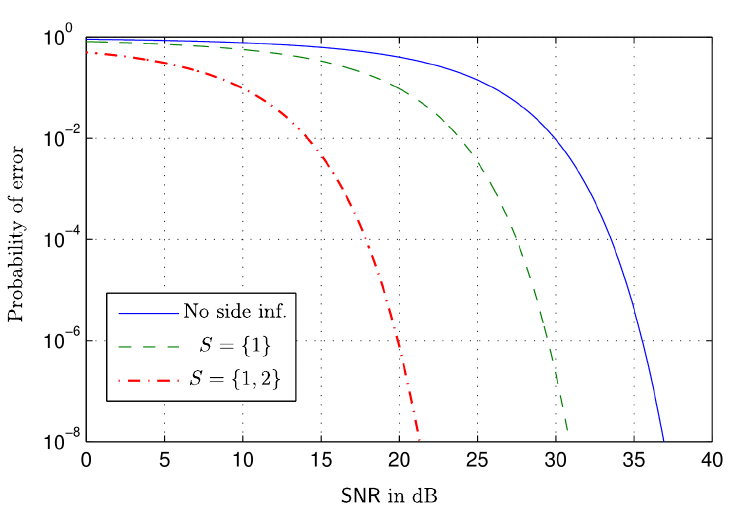}
\caption{Performance of the code of Example~\ref{ex:toy_example} for three different receivers.}
\label{fig:toy_example}
\end{figure*}

Consider \mbox{$K=3$} independent messages $w_1,w_2$ and $w_3$ assuming values from \mbox{$\mathcal{W}_1=\{0,1\}$}, \mbox{$\mathcal{W}_2=\{0,1,2\}$} and \mbox{$\mathcal{W}_3=\{0,1,2,3,4\}$}, respectively. The three messages are encoded to a code \mbox{$\mathscr{C} \subset \Zb$} using the function
\begin{align*}
 x = 15w_1 + 10w_2 + 6w_3 {\rm~mod~} 30,
\end{align*}
where the operation $a {\rm~mod~} 30$ gives the unique remainder in $\mathscr{C}=\{-15,-14,\dots,13,14\}$ when the integer $a$ is divided by $30$. Using Chinese remainder theorem~\cite{Ros_Pearson_05}, it is easy to verify that $\mathscr{C}$ is the set of all possible values that the transmit symbol $x$ can assume.
Since the dimension of $\mathscr{C}$ is \mbox{$n=1$}, the rate of the $k^{\text{th}}$ message is $R_k=\log_2|\mathcal{W}_k|$~b/dim, i.e.,
\begin{align*}
R_1 = 1, \, R_2 = \log_2 3, \text{ and } R_3 = \log_2 5 \text{ b/dim}.
\end{align*}

With no side information, a receiver decodes the channel output to the nearest point in $\mathscr{C}$, with the corresponding minimum inter-codeword distance \mbox{$d_0=1$}.
%
With \mbox{$S=\{1\}$}, the receiver knows the value of the first message \mbox{$w_1=a_1$}. The decoder of this receiver restricts the choice of transmit symbols to the subcode
\begin{align*}
\mathscr{C}_{a_1}=\left\{ 15a_1 + 10w_2 + 6w_3 {\rm~mod~} 30 \, \vert \, w_2 \in \mathcal{W}_2, w_3 \in \mathcal{W}_3 \right\}.
\end{align*}
Any two points in this subcode differ by \mbox{$10 \Delta w_2 + 6 \Delta w_3$}, where $\Delta w_2$ and $\Delta w_3$ are integers, not both equal to zero. Since the greatest common divisor ($\gcd$) of $10$ and $6$ is \mbox{$\gcd(10,6)=2$}, the minimum non-zero magnitude of \mbox{$10\Delta w_2 + 6 \Delta w_3$} is $2$~\cite{Ros_Pearson_05}. Hence, the minimum distance corresponding to the side information index set \mbox{$S=\{1\}$} is \mbox{$d_S=2$}. The side information rate is \mbox{$R_S=R_1=1$}~b/dim, which equals \mbox{$\log_2 d_S$}.

When \mbox{$S=\{1,2\}$}, the set of possible transmit symbols is
\begin{align*}
\mathscr{C}_{(a_1,a_2)}=\left\{15a_1 + 10a_2 + 6w_3 {\rm~mod~} 30 |w_3 \in \mathcal{W}_3\right\},
\end{align*}
where $w_1=a_1$ and $w_2=a_2$ are known. The minimum distance of this subcode is \mbox{$d_S=6$}, and the side information rate is \mbox{$R_S = R_1 + R_2 = \log_2 6 = \log_2 d_S$}~b/dim.

Similarly, for every choice of {$S \subset \{1,2,3\}$}, we have \mbox{$R_S = \log_2 d_S$}, i.e., the minimum distance $d_S$ is exponential in the side information rate $R_S$.
As will be shown in Sections~\ref{sec:commutative} and~\ref{sec:Hurwitz}, this property is satisfied by all the proposed lattice index codes.
Using {$R_S=\log_2 d_S$} in~\eqref{eq:side_inf_gain}, we see that the side information gain is uniform, and \mbox{$\Gamma=20 \log_{10} 2 \approx 6\text{ dB/b/dim}$}.
In Section~\ref{sec:upper_bound} we show that this is the maximum side information gain achievable by any index code \mbox{$\mathscr{C} \subset \Zb$} in which the messages are linearly encoded.
Fig.~\ref{fig:toy_example} shows the performance of the code with \mbox{$S=\varnothing$}, \mbox{$S=\{1\}$} and \mbox{$S=\{1,2\}$}. At the probability of error of $10^{-4}$, the side informations corresponding to \mbox{$S=\{1\}$} and \mbox{$S=\{1,2\}$} provide ${\sf SNR}$ gains of $6$~dB and $15.6$~dB over \mbox{$S=\varnothing$}. This is close to the corresponding squared distance gains of $10 \log_{10} \left( 2^2 \right)$~dB and $10 \log_{10} \left( 6^2 \right)$~dB, respectively. \hfill\IEEEQED
\end{example}

\begin{figure*}[t!]
\centering
\subfloat[The $16$-PSK index code represented as a labelling scheme.]{\includegraphics[width=3in]{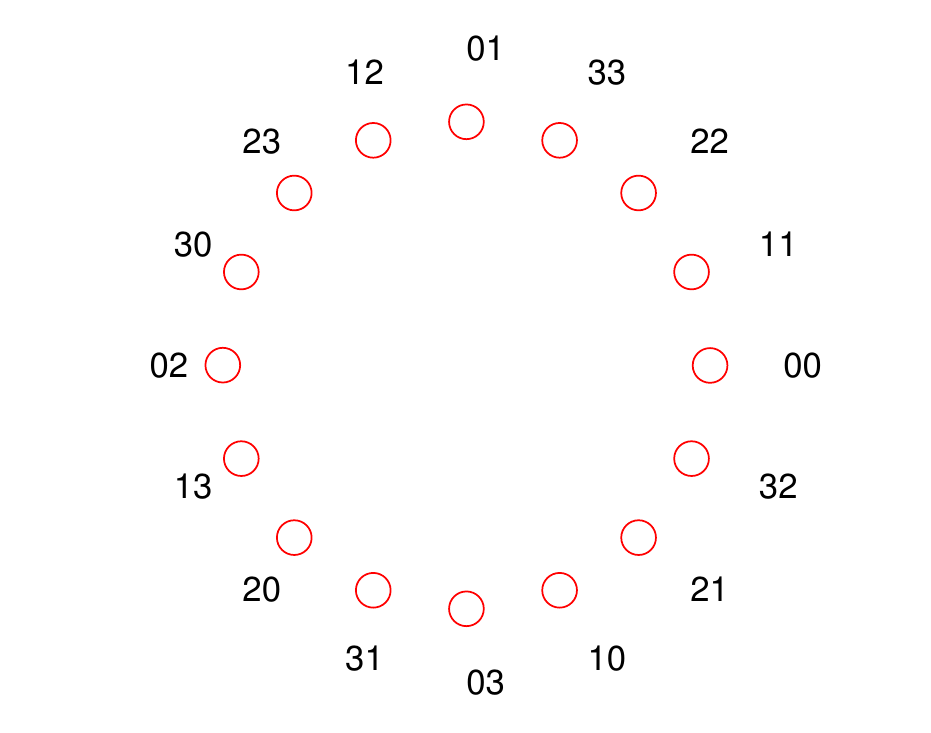} \label{fig:16PSK_1}}
\\
\subfloat[The filled circles denote the codewords corresponding to \mbox{$w_1=0$}.]{\includegraphics[width=3in]{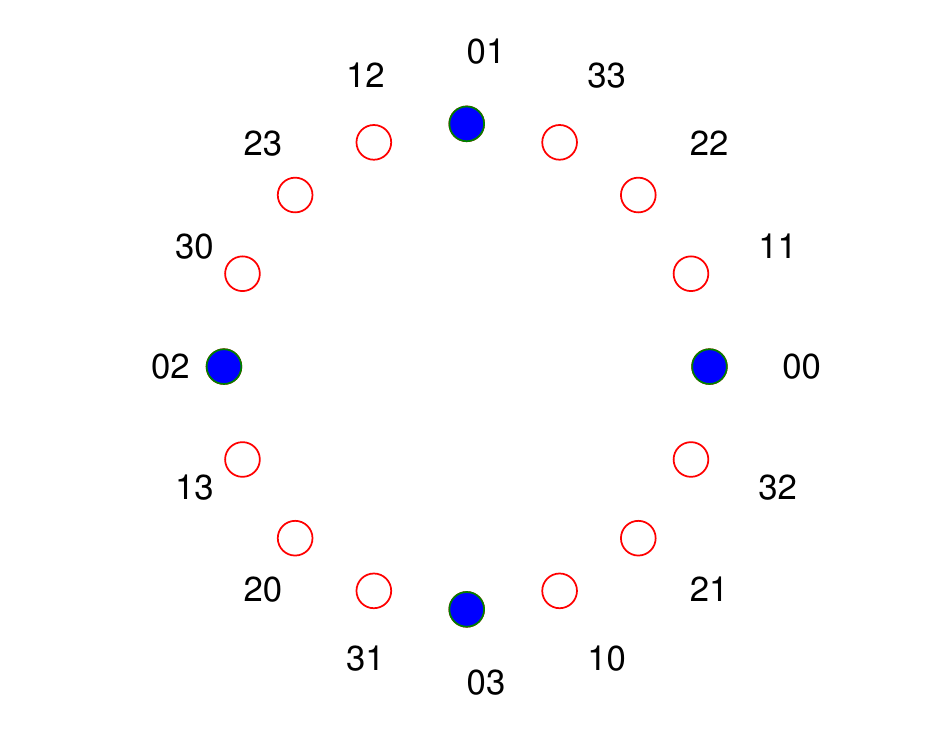} \label{fig:16PSK_2}}
\hfil
\subfloat[The filled circles denote the codewords corresponding to \mbox{$w_2=0$}.]{\includegraphics[width=3in]{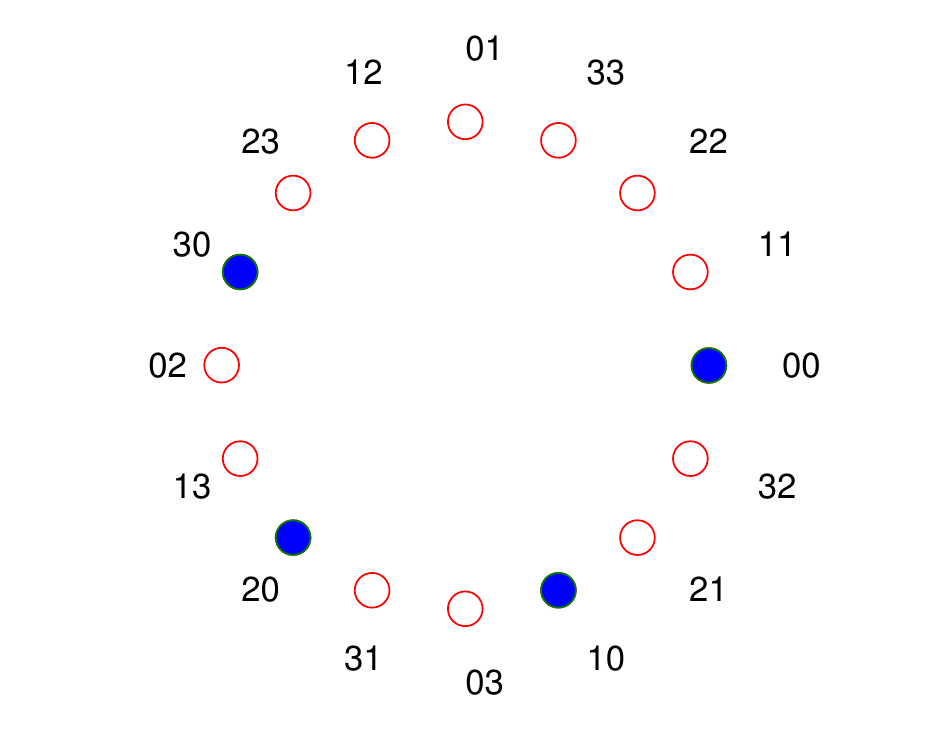} \label{fig:16PSK_3}}
\caption{The $16$-PSK index code of Example~\ref{ex:16PSK} that encodes two $4$-ary messages and provides $\Gamma=9.1$~dB/b/dim.}
\end{figure*}
We now give an example of a non-uniform gain index code with \mbox{$\Gamma > 20 \log_{10} 2$}~dB/b/dim based on a non-lattice constellation.
This example also highlights the notion that, given a constellation $\mathscr{C}$, the task of designing a good index code is equivalent to designing a good labelling scheme.
\begin{example}[\emph{A 2-message index code using $16$-PSK}] \label{ex:16PSK}
We encode $K=2$ messages with alphabets $\mathcal{W}_1=\mathcal{W}_2=\{0,1,2,3\}$ to the $16$-PSK constellation $\mathscr{C}$.
The encoder $\rho: \mathcal{W}_1 \times \mathcal{W}_2 \to \mathscr{C}$ is represented as a labelling scheme in Fig.~\ref{fig:16PSK_1} where each of the $16$ constellation points $x$ is labelled with the corresponding message tuple $(w_1,w_2)=\rho^{-1}(x)$.
The dimension of the code is $n=2$, and the message rates are
\begin{equation*}
R_1=R_2=\frac{1}{2} \log_2 4 = 1 \text{ b/dim}.
\end{equation*}

A receiver with no side information, i.e., with \mbox{$S=\varnothing$}, decodes the received channel vector to the nearest $16$-PSK constellation point. The error performance at this receiver is equal to that of the $16$-PSK signal set.
Assuming that the constellation points have unit energy, the corresponding minimum Euclidean distance at this receiver is $d_0=2\sin\left(\sfrac{\pi}{16}\right)$.

If \mbox{$S=\{1\}$}, the receiver has the knowledge of the value of the first message $w_1$. For example, if $w_1=0$, this receiver knows that the transmitted vector is one of the four points in the set $\{\rho(0,w_2) \, \vert \, w_2 \in \mathcal{W}_2 \}$; see Fig.~\ref{fig:16PSK_2}. The minimum Euclidean distance of this subcode is $2\sin\left(\sfrac{\pi}{4}\right)=\sqrt{2}$. The minimum Euclidean distance corresponding to the other three values of $w_1$ is also $\sqrt{2}$. Hence, for \mbox{$S=\{1\}$}, we have \mbox{$d_S=\sqrt{2}$} and the normalized squared distance gain is $10\log_{10}\left(\sfrac{d_S^2}{d_0^2}\right)/R_S=11.2$~dB/b/dim.

A receiver with \mbox{$S=\{2\}$} decodes its channel output to one of the four subcodes of $\mathscr{C}$ determined by the value of $w_2$ obtained as side information. The subcode for \mbox{$w_2=0$} is shown in Fig.~\ref{fig:16PSK_3}. All four subcodes have minimum Euclidean distance \mbox{$d_S=2\sin\left(\sfrac{3\pi}{16}\right)$}. The squared distance gain for $S=\{2\}$ normalized by $R_S$ is $9.1$~dB/b/dim. To conclude, this $16$-PSK index code does not have uniform gain, and has \mbox{$\Gamma=\min\{11.2,9.1\}=9.1$~dB/b/dim}.
\hfill\IEEEQED
\end{example}

\begin{figure*}[!t]
\centering
\includegraphics[width=4in]{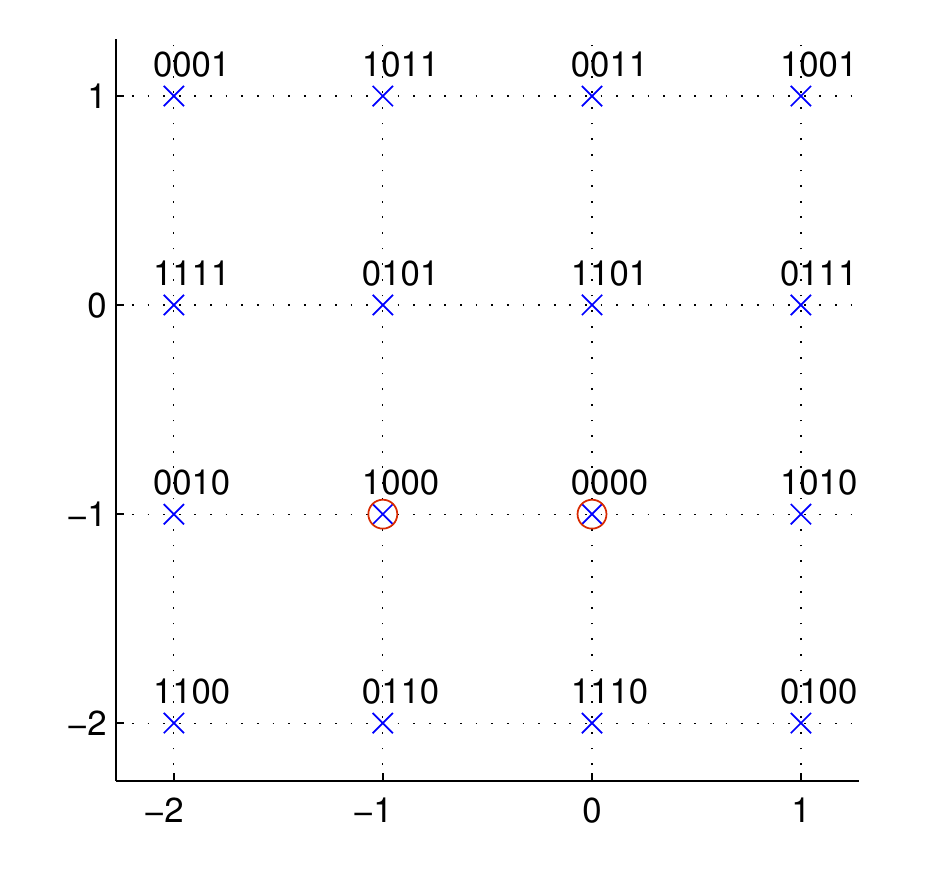}
\caption{A set partition labelling of $16$-QAM. The two points marked with circles form the subcode for the side information \mbox{$(w_2,w_3,w_4)=(0,0,0)$}.}
\label{fig:set_part_constellation}
\end{figure*}

\begin{figure*}[!t]
\centering
\includegraphics[width=5in]{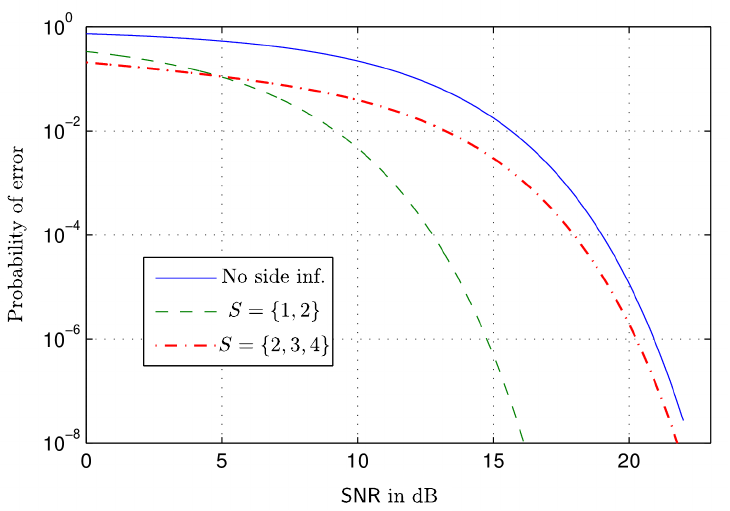}
\caption{Performance of set partition labelling of Example~\ref{ex:set_partition}.}
\label{fig:set_part_simulation}
\end{figure*}

\begin{example}[\emph{A bad index code}] \label{ex:set_partition}
Labelling a given constellation $\mathscr{C}$ by \emph{set partitioning}~\cite{Ung_IEEE_IT_82} is apparently a related problem, but it does not necessarily provide good index codes. In set partitioning with binary `labels' $w_1,\dots,w_K$, the constellation $\mathscr{C}$ is recursively partitioned into two smaller signal sets with larger minimum distance. For any $S=\{1,2,\dots,k\}$, \mbox{$k < K$}, the set of points with a given label \mbox{$w_S=a_S$} forms one of the $2^k$ $k^{\text{th}}$-level partitions of $\mathscr{C}$. The minimum distance of the partition improves with increasing $k$. Fig.~\ref{fig:set_part_constellation} shows one such labelling of $16$-QAM, with \mbox{$K=4$}, where the knowledge of the values of the first $k$ bits $w_1,\dots,w_k$ increases the minimum distance from \mbox{$d_0=1$} to \mbox{$d_S=\sqrt{2^k}$}. However, this does not guarantee squared distance gain for every side information index set \mbox{$S \subset \{1,\dots,K\}$}. For instance, the side information \mbox{$(w_2,w_3,w_4)=(0,0,0)$}, corresponding to \mbox{$S=\{2,3,4\}$}, does not provide any improvement in minimum distance. The performance of the code of Fig.~\ref{fig:set_part_constellation} for \mbox{$S=\varnothing$}, \mbox{$S=\{1,2\}$} and \mbox{$S=\{2,3,4\}$} is shown in Fig.~\ref{fig:set_part_simulation}. When the error rate is \mbox{$P_e=10^{-4}$}, the knowledge of the first two bits provides an ${\sf SNR}$ gain of $6.2$~dB. However, the ${\sf SNR}$ gain with \mbox{$S=\{2,3,4\}$} is only $1$~dB at \mbox{$P_e=10^{-4}$} and is smaller for diminishing $P_e$. \hfill\IEEEQED
\end{example}

Set partition labelling is designed to provide squared distance gain when $S$ is of the form \mbox{$\{1,2,\dots,k\}$} for \mbox{$k<K$}. When restricted to such side information index sets, set partitioning provides side information gain $\sim 6$~dB/b/dim. The codes in Examples~\ref{ex:toy_example} and~\ref{ex:16PSK} allow us to achieve side information gains when $S$ is any subset of $\{1,\dots,K\}$.

\section{Lattice Index Codes} \label{sec:linear_lattice_codes}

We first review the necessary background on lattices and lattice codes, based on~\cite{CoS_Springer_99,For_IEEE_IT_88,ZSE_IEEE_IT_02}~(Section~\ref{sec:lattices}), introduce lattice index codes~(Section~\ref{sec:linear_modulo_encoding}), and then derive an upper bound on the side information gain of such codes constructed from the densest lattices~(Section~\ref{sec:upper_bound}).

\subsection{Lattices and lattice codes} \label{sec:lattices}

An $n$-dimensional \emph{lattice} in $\Rb^n$ is a discrete additive subgroup \mbox{$\Lambda=\{Gz \, | \, z \in \Zb^n\}$}, where the full-ranked matrix \mbox{$G \in \Rb^{n \times n}$} is called the \emph{generator matrix} of $\Lambda$. Since the difference between any two lattice points is also a lattice point, the \emph{minimum distance} $d_{\min}\left(\Lambda\right)$ between any two points in $\Lambda$ is the Euclidean length of the shortest non-zero vector of $\Lambda$. The \emph{closest lattice point quantizer} \mbox{$Q_{\Lambda}:\Rb^n \to \Lambda$} is
\begin{align*}
Q_{\Lambda}(x) = \lambda \text{ if } \|x -\lambda \| \leq \| x - \lambda' \| \text{ for every } \lambda' \in \Lambda,
\end{align*}
where \mbox{$x \in \Rb^n$}, \mbox{$\lambda \in \Lambda$}, and ties (if any) between competing lattice points are broken systematically. The \emph{fundamental Voronoi region} $\mathcal{V}_{\La}$ is the set of all points in $\Rb^n$ that are mapped to $0$ under $Q_{\Lambda}$. The volume of the fundamental region \mbox{$\vol(\Lambda)=\int_{\mathcal{V}_{\La}} \, dx$} is related to the generator matrix $G$ as \mbox{$\vol(\Lambda)=\vert \det G \vert$}.
The \emph{packing radius} \mbox{$r_{\rm pack}(\La)=\sfrac{d_{\min}(\Lambda)}{2}$} is the radius of the largest $n$-dimensional sphere contained in the Voronoi region $\mathcal{V}_{\La}$.
The \emph{center density} of $\La$ is
\begin{align} \label{eq:delta}
 \delta(\La) = \frac{\left( r_{\rm pack}(\La) \right)^n}{\vol(\La)} =  \frac{\left( \sfrac{d_{\min}(\La)}{2} \right)^n}{\vol(\La)}.
\end{align}
The center density of a lattice is invariant to scaling, i.e., \mbox{$\delta(\La)=\delta(\alpha\La)$} for any non-zero \mbox{$\alpha \in \Rb$}.
If $\La$ is scaled by \mbox{$\alpha=\sfrac{2}{d_{\min}(\La)}$}, then \mbox{$r_{\rm pack}\left(\alpha\La\right)=1$} and \mbox{$\delta=\sfrac{1}{\vol(\alpha\La)}$} is the average number of points in $\alpha\La$ per unit volume in $\Rb^n$, 
i.e., $\delta$ is the density of the lattice points in $\Rb^n$ when scaled to unit packing radius.
For the same average transmit power constraint and minimum distance, a constellation carved from a lattice with a higher value of $\delta$ has a larger size, and hence, a higher coding gain. The densest lattices are known for dimensions \mbox{$n=1,2,\dots,8$} and \mbox{$n=24$}~\cite{CoS_Springer_99,CoK_AnnMath_04}.
For \mbox{$n=1,\dots,8$}, the densest lattices are $\Zb,A_2,D_3,D_4,D_5,E_6,E_7$ and $E_8$, respectively, while the Leech lattice $\La_{24}$ is densest in $24$ dimensions. The lattice $D_4$ is equivalent to its dual lattice $D_4^*$ up to scaling and orthogonal transformation. Hence, $D_4^*$ too has the highest density in $4$ dimensions.

The \emph{modulo}-$\Lambda$ operation
\mbox{$x {\rm~mod~} \La = x - Q_{\La}(x) \in \mathcal{V}_{\La}$},
is the difference between a vector and its closest lattice point, and it satisfies the relation
\begin{align} \label{eq:modulo_distributive}
{ \left( x_1 + x_2 \right) {\rm~mod~} \La = \left( x_1 {\rm~mod~} \La + x_2  \right) {\rm~mod~} \La}
\end{align}
for all \mbox{$x_1,x_2 \in \Rb^n$}.
Let \mbox{$\Ls \subset \La$} be a sub-lattice of $\La$, and $\Lambda/\Ls$ be the quotient group of the cosets of $\Ls$ in $\La$. Each coset of $\La/\Ls$ can be identified by its representative contained in $\mathcal{V}_{\Ls}$. We will identify the group $\La/\Ls$ with the group of coset leaders \mbox{$\La \cap \mathcal{V}_{\Ls}=\La {\rm~mod~} \Ls$}, where addition is performed modulo $\Ls$. Further,
\begin{align*}
\mbox{$|\La/\Ls| = |\La {\rm~mod~} \Ls| = \frac{\vol(\Ls)}{\vol(\La)}$}.
\end{align*}
The constellation $\La/\Ls$ is called a \emph{(nested) lattice code}, and $\Ls$ is called the \emph{coarse lattice} or the \emph{shaping lattice}~\cite{For_IEEE_IT_88,ZSE_IEEE_IT_02}.

\subsection{Lattice index codes} \label{sec:linear_modulo_encoding}

Consider $K$ lattices \mbox{$\La_1,\dots,\La_K$}, with a common sub-lattice \mbox{$\Ls \subset \La_k$}, \mbox{$k=1,\dots,K$}. We will use the lattice constellations \mbox{$\La_1/\Ls,\dots,\La_K/\Ls$} as the alphabets \mbox{$\mathcal{W}_1,\dots,\mathcal{W}_K$} of the $K$ messages at the source.

\begin{definition} \label{def:lattice_index_code}
A \emph{lattice index code} for $K$ messages consists of $K$ lattice constellations $\La_1/\Ls,\dots,\La_K/\Ls$, and the injective linear encoder map $\rho:\La_1/\Ls \times \cdots \times \La_K/\Ls \to \mathscr{C}$ given by
\begin{align} \label{eq:rho}
\rho\left(x_1,\dots,x_K\right) = \left( x_1 + \cdots + x_K \right) {\rm~mod~} \Ls,
\end{align}
where $x_k \in \La_k/\Ls$ and $\mathscr{C}$ is the set of all possible values of the transmit symbol $x=\rho(x_1,\dots,x_K)$. \hfill\IEEEQED
\end{definition}

We require that $\rho$ be injective so that no two message tuples are mapped to the same transmit symbol.
We now relate some properties of a lattice index code to those of its component lattice constellations $\La_1/\Ls,\dots,\La_K/\Ls$.

\begin{itemize}
\item \emph{The transmit codebook $\mathscr{C}$:} Let \mbox{$\La=\La_1+\cdots+\La_K$} be the lattice generated by the union of the lattices {$\La_1,\dots,\La_K$}.
It follows from~\eqref{eq:rho} that \mbox{$x_1+\dots+x_K \in \La$}, and hence $x \in \La/\Ls$.
On the other hand, every point in $\La$ is the sum of $K$ lattice points, one each from \mbox{$\La_1,\dots,\La_K$}. It follows from~\eqref{eq:modulo_distributive} that every point in the lattice constellation $\La/\Ls$ is the ${\rm mod~}\Ls$ sum of $K$ points, from \mbox{$\La_1/\Ls,\dots,\La_K/\Ls$}, respectively. Hence, the transmit codebook is \mbox{$\mathscr{C}=\La/\Ls$}.

\item \emph{Message rates:} If $\La$ is an $n$-dimensional lattice, the rate of the $k^{\text{th}}$ message is
\begin{align*}
R_k = \frac{1}{n} \log_2 |\mathcal{W}_k| &= \frac{1}{n} \log_2 |\La_k/\Ls|\\ &= \frac{1}{n} \log_2 \frac{\vol(\Ls)}{\vol(\La_k)} \text{ b/dim}.
\end{align*}

\item \emph{Minimum distance:}  Since \mbox{$\mathscr{C}=\La/\Ls$} is carved from the lattice $\La$, the minimum inter-codeword distance with no side information is
\begin{align} \label{eq:d0}
d_0=d_{\min}(\La).
\end{align}
Now suppose that a receiver has side information of the messages with indices in $S$, say $x_S=a_S$ (i.e., $x_k=a_k$, $k \in S$).
The subcode $\mathscr{C}_{a_S}$ decoded by the receiver is 
\begin{align*}
&~~~\left\{ \sum_{k \in S} a_k + \sum_{k \in S^{\mathsf{c}}} x_k  \Big\vert  x_k \in \La_k/\Ls, k \in S^{\mathsf{c}}  \right\} {\rm mod~} \Ls \\
&= \left( \sum_{k \in S} a_k + \sum_{k \in S^{\mathsf{c}}} \La_k/\Ls \right) {\rm~mod~} \Ls \\
&= \left( \sum_{k \in S} a_k + \sum_{k \in S^{\mathsf{c}}} \La_k  \right) {\rm~mod~\Ls},
\end{align*}
where we have used~\eqref{eq:modulo_distributive}.  
Thus, $\mathscr{C}_{a_S}$ is a lattice code carved from a translate of the lattice $\sum_{k \in S^{\mathsf{c}}}\La_k$, and hence its minimum distance is 
\begin{align} \label{eq:dS_lambda}
d_S = d_{\min}\left( \sum_{k \in S^{\mathsf{c}}} \La_k \right).
\end{align}

\end{itemize}

\begin{example}
The code in Example~\ref{ex:toy_example} is a lattice index code with \mbox{$K=3$}, \mbox{$\La_1=15\Zb$}, \mbox{$\La_2=10\Zb$}, \mbox{$\La_3=6\Zb$}, \mbox{$\Ls=30\Zb$} and \mbox{$\La=15\Zb+10\Zb+6\Zb=\Zb$}.\hfill\IEEEQED
\end{example}

The transmit codebook \mbox{$\mathscr{C}=\La/\Ls$} of a lattice index code is a commutative group under addition modulo $\Ls$, and $\La_1/\Ls,\dots,\La_K/\Ls$ are subgroups of $\mathscr{C}$. It follows from Definition~\ref{def:lattice_index_code} that the encoding map $\rho$ is a group isomorphism between $\mathscr{C}$ and the direct product $\La_1/\Ls \times \cdots \times \La_K/\Ls$ of the subgroups $\La_1/\Ls,\dots,\La_K/\Ls$, i.e., $\mathscr{C}$ is a direct sum of these $K$ subgroups.
Thus, the problem of designing a good lattice index code is to construct a pair \mbox{$\Ls \subset \La$} of nested lattices, and to find a decomposition of $\La/\Ls$ into $K$ subgroups, such that \mbox{$d_S=d_{\min}\left( \sum_{k \in S^{\mathsf{c}}} \La_k \right)$} is large for every choice of \mbox{$S \subset \{1,\dots,K\}$}.
While constructions of pairs \mbox{$\Ls \subset \La$} of lattices~\cite{For_IEEE_IT_88,ZSE_IEEE_IT_02} and chains \mbox{$\La \subset \La' \subset \La'' \subset \cdots$} of nested lattices~\cite{ZSE_IEEE_IT_02} are well known in the literature, we require a lattice code $\La/\Ls$ and a set of its generating subcodes \mbox{$\La_1/\Ls,\dots,\La_K/\Ls$} such that all non-trivial direct sums $\sum_{k \in S^{\mathsf{c}}} \La_k/\Ls$, \mbox{$S \subset \{1,\dots,K\}$}, of the $K$ subcodes have large minimum Euclidean distances.

In Sections~\ref{sec:commutative} and~\ref{sec:Hurwitz}, we construct index codes using lattices that possess the multiplicative structure of a \emph{principal ideal domain} (PID) or that of a \emph{module} over a PID, besides the additive structure of a commutative group. The structure of a PID (or a module over a PID) enables us to control the minimum Euclidean distance $d_S$, and hence the side information gain $\Gamma$, of the resulting codes.
When the underlying PID is commutative (Section~\ref{sec:commutative}), we use the Chinese remainder theorem to construct pairs \mbox{$\Ls \subset \La$} of nested lattices and decompose the resulting code $\La/\Ls$ into a direct sum of $K$ lattice subcodes.
We then construct lattice index codes using the Hurwitz integral quaternions as the base PID (Section~\ref{sec:Hurwitz}).
The Chinese remainder theorem does not apply to quaternions due to the technical reason that they are non-commutative and their ideals are not two-sided.
Nevertheless, we design a family of quaternionic lattice index codes by identifying the essential constituents of the techniques used in Section~\ref{sec:commutative} and extending them to the non-commutative case.

\subsection{An upper bound on the side information gain} \label{sec:upper_bound}

Consider the side information index set {$S = \{1,\dots,K-1\}$}. The minimum distance is
\begin{align*}
d_S = d_{\min}\left( \sum_{k \in S^{\mathsf{c}}}\La_k \right)= d_{\min}\left( \La_K \right),
\end{align*}
and the side information rate is
\begin{align*}
R_S &= R_{1} + \cdots + R_{K-1} = \frac{1}{n} \log_2 |\mathscr{C}| - R_K \\ 
    &= \frac{1}{n} \log_2 |\La/\Ls| - \frac{1}{n} \log_2 |\La_K/\Ls| \\
    &= \frac{1}{n} \log_2 \frac{\vol(\Ls)}{\vol(\La)} - \frac{1}{n} \log_2 \frac{\vol(\Ls)}{\vol(\La_K)} \\
    &= \frac{1}{n} \log_2 \frac{\vol(\La_K)}{\vol(\La)}.
\end{align*}
Representing the volume of the fundamental region in terms of the minimum distance $d_{\min}$ and the center density $\delta$ (see~\eqref{eq:delta}), 
\begin{align}
R_S &= \frac{1}{n} \log_2 \left(\frac{d_{\min}(\La_K)}{d_{\min}(\La)}\right)^n + \frac{1}{n} \log_2 \frac{\delta(\La)}{\delta(\La_K)} \nonumber \\
    &= \log_2 \frac{d_S}{d_0} + \frac{1}{n} \log_2 \frac{\delta(\La)}{\delta(\La_K)}, \label{eq:R_S_and_dmin}
\end{align}
If $\La$ is the densest lattice in $n$ dimensions, then \mbox{$\delta(\La) \geq \delta(\La_K)$}, and hence \mbox{$R_S \geq \log_2 \left( \sfrac{d_S}{d_0}  \right)$}. Thus the side information gain of $\mathscr{C}$ can be upper bounded as follows
\begin{align*}
\Gamma(\mathscr{C}) &= \min_S \frac{20 \log_{10} \left( \sfrac{d_S}{d_0} \right)}{R_S}
                    \leq \frac{20 \log_{10} \left( \sfrac{d_S}{d_0} \right)}{R_S} \\
                    &\leq \frac{20 \log_{10} \left( \sfrac{d_S}{d_0} \right)}{\log_2\left( \sfrac{d_S}{d_0} \right)}
                    = 20 \log_{10}2 \approx 6 \text{ dB/b/dim}.
\end{align*}

This upper bound on the side information gain holds only for the family of lattice index codes in which the underlying lattice $\La$ has the highest density in its dimension, such as when $\Lambda$ is $\Zb$, $A_2$ or $D_4^*$. This upper bound is independent of the information-theoretic result of~\cite{Tun_IEEE_IT_06} which guarantees the existence of codes that provide an ${\sf SNR}$ gain of \mbox{$\sim 6$~dB} for each b/dim of side information at the receiver.
The ${\sf SNR}$ gain of \mbox{$\sim 6$~dB/b/dim} of~\cite{Tun_IEEE_IT_06} holds for capacity-approaching noisy index codes at finite values of ${\sf SNR}$ in the asymptotic regime where the code dimension goes to infinity and the probability of error is arbitrarily small.
On the other hand, $\Gamma$ measures the squared distance gain at a finite code dimension, and approximates the ${\sf SNR}$ gain due to receiver side information in the high ${\sf SNR}$ regime.

When $\La$ is not the densest lattice in $\Rb^n$, for example when {$\La=\Zb^2$}, it is possible to have \mbox{$\delta(\La_K) > \delta(\La)$}. In such cases, from~\eqref{eq:R_S_and_dmin}, \mbox{$R_S < \log_2 \left( \sfrac{d_S}{d_0} \right)$}, and $\Gamma$ may exceed $\sim 6$~dB/b/dim.
Note that $\Gamma$ is a relative gain measured with respect to the performance of \mbox{$\mathscr{C}=\La/\Ls$} with no side information.
Any amount of side information gain available over and above \mbox{$\sim 6$~dB/b/dim} is due to the lower packing efficiency of $\La$ when compared to $\La_K$, and hence due to the inefficiency of $\mathscr{C}$ as a code in the point-to-point AWGN channel.
%
We now give an example of such a lattice index code with side information gain more than \mbox{$\sim 6$~dB/b/dim}.
\begin{example} \label{ex:lattice_more_than_6}
Consider \mbox{$K=2$} lattices $\La_1$ and $\La_2$ with generator matrices
\begin{equation} \label{eq:ex:G}
 G_1 = \begin{pmatrix} 4 & 2 \\ 0 & 3 \end{pmatrix} \text{ and } G_2 = \begin{pmatrix} 0 & 3 \\ 4 & 2 \end{pmatrix},
\end{equation}
respectively, and the coarse lattice \mbox{$\Ls=12\Zb^2$}.
The above lattices have been carefully chosen so that the densities of $\La_1$ and $\La_2$ are greater than that of their sum lattice $\La=\La_1+\La_2$.
In order to prove that this choice of $\La_1,\La_2$ and $\Ls$ indeed defines a valid lattice index code, we first show that $\Ls$ is a sub-lattice of $\La_1$ and $\La_2$, we then identify the transmit lattice $\La$ and the codebook $\mathscr{C}$, and then show that the encoding map $\rho$ is injective. Finally, we compute the minimum distances of $\La_1,\La_2$ and $\La$, and the side information gain $\Gamma$.

The following identities show that the basis vectors $\begin{pmatrix} 12,0 \end{pmatrix}^\intercal$ and $\begin{pmatrix} 0, 12 \end{pmatrix}^\intercal$ of \mbox{$\Ls=12\Zb^2$} can be expressed as integer linear combinations of the columns of $G_1$, and hence, $\Ls \subset \La_1$:
\begin{equation*}
\begin{pmatrix} 12 \\ 0 \end{pmatrix} = 3 \begin{pmatrix} 4 \\ 0 \end{pmatrix}, \text{ and }
\begin{pmatrix} 0 \\ 12 \end{pmatrix} =  - 2 \begin{pmatrix} 4 \\ 0 \end{pmatrix} + 4 \begin{pmatrix} 2 \\ 3 \end{pmatrix}.
\end{equation*}
Similarly, the proof for $\Ls \subset \La_2$ follows from the observation
\begin{equation*}
\begin{pmatrix} 12 \\ 0 \end{pmatrix} = -2 \begin{pmatrix} 0 \\ 4 \end{pmatrix} + 4 \begin{pmatrix} 3 \\ 2 \end{pmatrix}, \text{ and }
\begin{pmatrix} 0 \\ 12 \end{pmatrix} =  3 \begin{pmatrix} 0 \\ 4 \end{pmatrix}.
\end{equation*}
\end{example}

In order to identify the lattice \mbox{$\La=\La_1 + \La_2$}, we first note that \mbox{$\La_1,\La_2 \subset \Zb^2$}, and hence, \mbox{$\La \subset \Zb^2$}. The following expressions show that the basis vectors $\begin{pmatrix}1,0\end{pmatrix}^\intercal$ and $\begin{pmatrix}0,1\end{pmatrix}^\intercal$ of $\Zb^2$ are integer linear combinations of the columns of $G_1$ and $G_2$:
\begin{equation*}
\begin{pmatrix} 1 \\ 0 \end{pmatrix} = 2 \begin{pmatrix} 2 \\ 3 \end{pmatrix} - \begin{pmatrix} 0 \\ 4 \end{pmatrix} - \begin{pmatrix} 3 \\ 2 \end{pmatrix},
\begin{pmatrix} 0 \\ 1 \end{pmatrix} = 2 \begin{pmatrix} 3 \\ 2 \end{pmatrix} - \begin{pmatrix} 4 \\ 0 \end{pmatrix} - \begin{pmatrix} 2 \\ 3 \end{pmatrix}.
\end{equation*}
We conclude that $\La \supset \Zb^2$, and therefore, $\La = \Zb^2$. The transmit codebook $\mathscr{C}=\La/\Ls$ is $\Zb^2/12\Zb^2$. Thus, the encoding map $\rho$ has domain $\La_1/\Ls \times \La_2/\Ls$ and range $\mathscr{C}$. The cardinality of the domain is
\begin{align*}
|\La_1/\Ls| \cdot |\La_2/\Ls| = \frac{\vol(\Ls)}{\vol(\La_1)} \cdot \frac{\vol(\Ls)}{\vol(\La_2)} = \frac{144}{12} \cdot \frac{144}{12} = 144,
\end{align*}
and that of the range is
\begin{align*}
|\mathscr{C}| = |\La/\Ls| = \frac{\vol(\Ls)}{\vol(\La)} = \frac{144}{1} = 144.
\end{align*}
Since the domain and range are of the same cardinality, $\rho$ is injective, and consequently, $\mathscr{C}$ is a lattice index code. The dimension of this code is $n=2$, and the message rates are $R_1=R_2=\sfrac{1}{2} \log_2 12$~b/dim.

To calculate the side information gain of this code we require the values of $d_0$ and $d_S$, $S=\{1\},\{2\}$. From~\eqref{eq:d0}, $d_0=d_{\min}(\La)=d_{\min}(\Zb^2)=1$. From~\eqref{eq:dS_lambda}, $d_S=d_{\min}(\La_2)$ for $S=\{1\}$, and $d_S=d_{\min}(\La_1)$ for $S=\{2\}$. We now show that $d_{\min}(\La_1)=\sqrt{13}$. The proof for $d_{\min}(\La_2)=\sqrt{13}$ is similar.

From~\eqref{eq:ex:G}, we observe that every non-zero vector $x_1 \in \La_1$ is of the form $\begin{pmatrix} 4a+2b,3b\end{pmatrix}^\intercal$ for some $a,b \in \Zb$, both not equal to zero. The squared Euclidean length of $x_1$ is $$\|x_1\|^2=(4a+2b)^2+9b^2.$$ We now lower bound the value of $\|x_1\|^2$ based on the value of $b$. If $b=0$, $\|x_1\|^2=(4a)^2 \geq 16$. If $b$ is non-zero and even, we have $\|x_1\|^2 = (4a+2b)^2 + 9b^2 \geq 9b^2 \geq 9 \cdot 2^2=36$. When $b$ is non-zero and odd, we have $|2a+b| \geq 1$, and hence,
\begin{equation*}
\|x_1\|^2 = (4a+2b)^2 + 9b^2 = 4(2a+b)^2 + 9b^2 \geq 4 + 9b^2 \geq 13.
\end{equation*}
We conclude that $\|x_1\|^2 \geq 13$ for every non-zero \mbox{$x_1 \in \La_1$}. On the other hand, the choice of $a=0$, $b=1$ yields a vector $x_1$ with $\|x_1\|^2=13$. It follows that $d_{\min}(\La_1)=\sqrt{13}$.

The non-trivial subsets of \mbox{$\{1,\dots,K\}=\{1,2\}$} are \mbox{$S=\{1\}$} and \mbox{$S=\{2\}$}. For both these choices of $S$, we have
\begin{align*}
\frac{10 \log_{10}\left( \sfrac{d_S^2}{d_0^2}\right)}{R_S} = \frac{10 \log_{10} 13 }{\sfrac{1}{2} \log_2 12} &= 20\log_{10}2  \times \frac{\log_{10} 13}{\log_{10} 12} \\ &\approx 6.2\text{ dB/b/dim}.
\end{align*}
Since the normalized squared distance gain is the same for all choices of $S \subset \{1,\dots,K\}$, we conclude that $\mathscr{C}$ is a uniform gain lattice index code with $\Gamma \approx 6.2$~dB/b/dim.
The reason for $\Gamma$ to be more than $\sim 6$~dB/b/dim is that the lattices $\La_1$ and $\La_2$ have a larger center density than $\La$. For both $k=1,2$,
\begin{align*}
\delta(\La_k) = \frac{\left( \sfrac{d_{\min}(\La_k)}{2} \right)^n}{\vol(\La_k)} = \frac{\left(\sfrac{13}{2}\right)^2}{12} = \frac{13}{48},
\end{align*}
while $\delta(\La)=\delta(\Zb^2)=\sfrac{1}{4}$. \hfill\IEEEQED

\section{Construction of lattice index codes using Commutative PIDs} \label{sec:commutative}

In this section, we construct uniform gain index codes using lattices over commutative PIDs $\Zb$, $\Zb[i]$ and $\Zb[\omega]$ with \mbox{$\Gamma \approx 6$~dB/b/dim}. This includes the lattice $\Zb^2$, and the hexagonal lattice $A_2$ with generator matrix
\begin{align*}
 \begin{pmatrix} 1 & \sfrac{1}{2} \\[3pt] 0 & \sfrac{\sqrt{3}}{2} \end{pmatrix},
\end{align*}
which can be identified with $\Zb[i]$ and $\Zb[\omega]$, respectively.
%
In Section~\ref{sec:Hurwitz} we consider lattices over the Hurwitz integers which form a non-commutative PID.

\subsection{Review of commutative PIDs and complex lattices}

We assume that the reader is familiar with the notions of ideals and principal ideal domains. We now briefly recall some basic definitions and properties related to commutative PIDs and complex lattices. We refer the reader to~\cite{CoS_Springer_99} and~\cite{Jac_Freeman_74} for further details.

\subsubsection*{Commutative PIDs}

Let $\Db$ be a commutative ring with \mbox{$1 \neq 0$}. An \emph{ideal} $I$ in $\Db$ is an additive subgroup of $\Db$ with the property that \mbox{$ab \in I$} for every \mbox{$a \in I$} and \mbox{$b \in \Db$}. The ideal \emph{generated} by an element $a$ is the smallest ideal containing $a$, and is given by \mbox{$a\Db=\{ab\,\vert\,b\in\Db\}$}.
An ideal $I$ is \emph{principal} if it is generated by a single element of $\Db$, i.e., $I=a\Db$ for some $a \in \Db$.
If the product of any two non-zero elements of $\Db$ is non-zero, $\Db$ is said to be an \emph{integral domain}. If every ideal of an integral domain $\Db$ is principal, then $\Db$ is a \emph{principal ideal domain}~(PID). In the rest of this section we will assume that $\Db$ is a commutative PID.

For \mbox{$a,b \in \Db$} we say that $a$ is a divisor of $b$, i.e., \mbox{$a \divides b$} if \mbox{$b=da$} for some \mbox{$d \in \Db$}.
The \emph{units} of $\Db$ are the divisors of $1$, i.e., they are the elements with a multiplicative inverse.
Two elements \mbox{$a,b \in \Db$} are \emph{associates} if \mbox{$a=ub$} (or equivalently, \mbox{$b=u^{-1}a$}) for some unit $u$.

The $\gcd$ of $a$ and $b$ is the generator of the smallest ideal containing $a$ and $b$, i.e., \mbox{$a\Db + b\Db = \gcd(a,b)\Db$}.
The $\gcd$ is unique up to multiplication by a unit.
If $d \divides a$ and $d \divides b$, then $d \divides \gcd(a,b)$.
Two elements $a$ and $b$ are \emph{relatively prime} if \mbox{$\gcd(a,b)$} is a unit.
A non-unit \mbox{$\phi \in \Db$} is \emph{prime} if \mbox{$\phi \divides ab$} implies that either \mbox{$\phi \divides a$} or \mbox{$\phi \divides b$}. A prime can not be expressed as a product of two non-units. Any two non-associate primes are relatively prime.
Every PID is a \emph{unique factorization domain}, i.e., every non-zero element of $\Db$ can be factored as a product of primes, uniquely up to multiplication by units. If \mbox{$a=\phi_1^{e_1}\cdots\phi_K^{e_K}$} is the factorization of $a$ as a product of non-associate primes \mbox{$\phi_1,\dots,\phi_K$}, and \mbox{$d \divides a$}, then \mbox{$d=u \phi_1^{e_1'}\cdots\phi_K^{e_K'}$}, where $u$ is a unit and \mbox{$e_k' \leq e_k$} for \mbox{$k=1,\dots,K$}.

\subsubsection*{Complex lattices} Let $\Db$ be either $\Zb[i]$ or $\Zb[\omega]$. A $\Db$-\emph{lattice} $\Lt$ is a discrete subgroup of a complex Euclidean space that is closed under multiplication by elements \mbox{$m \in \Db$}. Since every $\Db$-lattice is isomorphic to a real lattice of twice its dimension, we will denote its complex dimension by $\nbytwo$, where the even integer $n$ is the real dimension.
Let $$\Lt=\left\{ \widetilde{G}z \, | \, z \in \Db^{\nbytwo} \right\}$$ be a $\Db$-lattice with the full-rank generator matrix \mbox{$\widetilde{G} \in \Cb^{\nbytwo \times \nbytwo}$}.
Let \mbox{$\Psi:\Cb^{\nbytwo} \to \Rb^n$} be the isomorphism that maps the complex vector $(v_1,\dots,v_{\nbytwo})^\intercal$ to the real vector
\begin{align*}
\left( \real(v_1),\dots,\real(v_{\nbytwo}),\imag(v_1),\dots,\imag(v_{\nbytwo}) \right)^\intercal.
\end{align*}
The real lattice associated with $\Lt$ is
\begin{align*}
\La = \Psi\left(\Lt\right) = \left\{ \Psi(v) \, \vert \, v \in \Lt \right\} \subset \Rb^n.
\end{align*}
The lattice $\La$ is called \emph{Gaussian} if $\Db=\Zb[i]$, and \emph{Eisenstein} if $\Db=\Zb[\omega]$.
The hexagonal lattice $A_2$, the root lattice $E_6$, and the Coxeter-Todd lattice $K_{12}$ can be viewed as Eisenstein lattices, while the checkerboard lattice $D_4$, the Gosset lattice $E_8$, the laminated lattices $\La_{12}^{\max}$, $\La_{16}$, and the Leech lattice $\La_{24}$ can be viewed as both Gaussian and Eisenstein lattices.
If $\Db=\Zb[i]$, the real generator matrix $G$ of $\La$ is related to the complex generator matrix $\widetilde{G}$ as
\begin{align} \label{eq:generator_Gaussian}
 G = \begin{pmatrix}[r]
     \real(\widetilde{G}) & -\imag(\widetilde{G}) \\[1mm]
     \imag(\widetilde{G}) &  \real(\widetilde{G})
     \end{pmatrix},
\end{align}
and if $\Db=\Zb[\omega]$,
\begin{align*}
 G = \begin{pmatrix}[r]
     \real(\widetilde{G}) & \sfrac{1}{2} \left( \real(\widetilde{G}) + \sqrt{3}\imag(\widetilde{G}) \right) \\[3mm]
     \imag(\widetilde{G}) & \sfrac{1}{2} \left( \imag(\widetilde{G}) - \sqrt{3}\real(\widetilde{G}) \right)
     \end{pmatrix}.
\end{align*}
Since $\Psi$ preserves addition, for any two complex lattices $\Lt_1$, $\Lt_2$, we have $$\Psi(\Lt_1 + \Lt_2)=\Psi(\Lt_1) + \Psi(\Lt_2).$$ Also, $\Lt_1 \subset \Lt_2$ if and only if $\Psi(\Lt_1) \subset \Psi(\Lt_2)$.

\vspace{2pt}
We will use the symbols $\vol(\Lt)$ and $d_{\min}(\Lt)$ to denote the volume and the length of the shortest vector of the associated real lattice $\La$, i.e.,
\begin{align*}
\vol\left(\Lt\right) \triangleq \vol\left(\Psi(\Lt)\right) \text{ and } d_{\min}\left(\Lt\right) \triangleq d_{\min}\left(\Psi(\Lt)\right).
\end{align*}

For both Gaussian and Eisenstein lattices, scaling $\Lt$ by a complex number $m \in \Cb$ is equivalent to left-multiplying the real generator matrix $G$ by
\begin{align*}
 \mathcal{M}(m) = \begin{pmatrix}[r]
                  \real(m) \, \mathbb{I} & -\imag(m) \, \mathbb{I} \\
                  \imag(m) \, \mathbb{I} &  \real(m)\, \mathbb{I}
                  \end{pmatrix},
\end{align*}
where $\mathbb{I}$ is the identity matrix of dimension $\nbytwo \times \nbytwo$. Observing that $\mathcal{M}(m)$ is an orthogonal matrix with determinant $|m|^n$, we have
\begin{align}
 \vol(m\Lt) &= |\det \mathcal{M}(m)| \cdot |\det G |  = |m|^n \, \vol(\La), \text{ and } \label{eq:vol_complex_lattice} \\
 d_{\min}(m\Lt) &= |m| \, d_{\min}(\La). \label{eq:dmin_complex_lattice}
\end{align}

\subsection{Construction of index codes using commutative PIDs} \label{sec:construction_comm_pid}

Let \mbox{$\Db \subset \Cb$} be a commutative PID. Consider $K$ non-associate primes \mbox{$\phi_1,\dots,\phi_K \in \Db$}, and their product \mbox{$M=\prod_{k=1}^{K}\phi_k$}. The Chinese remainder theorem~\cite[page~159]{Ros_Pearson_05} states that the direct product $\Db/\phi_1\Db \times \cdots \times \Db/\phi_K\Db$ is isomorphic to the quotient ring $\Db/M\Db$. The one-to-one correspondence between them is obtained using the map
\begin{align*}
(w_1,\dots,w_K) \to w_1M_1 + w_2 M_2 + \cdots + w_KM_K {\rm~mod~} M\Db,
\end{align*}
where \mbox{$w_k \in \Db/\phi_k\Db$} and \mbox{$M_k=\sfrac{M}{\phi_k}$}. Since $w_kM_k$ is an element of $M_k\Db/M\Db$, we observe that encoding the $K$ source messages individually using the constellations $M_1\Db/M\Db,\dots,M_K\Db/M\Db$, and generating the transmit symbol as their modulo-$M\Db$ sum gives an injective encoding map.
Further, given the side information $w_S=a_S$, corresponding to the index set $S \subset \{1,\dots,K\}$, the minimum distance $d_S$ between the valid codewords can be readily obtained as the magnitude of $\gcd(M_k,k \in S^{\mathsf{c}})$ (cf. Example~\ref{ex:toy_example}).
The codebook $\Db/M\Db$ can be thought of as a lattice index code built over the one-dimensional $\Db$-lattice \mbox{$\Lt=\Db$}.
In this section, we apply this encoding technique to arbitrary $\Db$-lattices and show that the resulting lattice index codes provide large side information gains.

We first describe our construction with complex lattices, i.e., \mbox{$\Db=\Zb[i]$} and $\Zb[\omega]$, and prove that it provides a uniform side information gain $\Gamma \approx 6$~dB/b/dim. We then briefly describe the case \mbox{$\Db=\Zb$}, the proof of which follows from simple modifications of the proofs of Lemmas~\ref{lem:arbit_construction} and~\ref{lem:dS_RS} below.

\subsection*{Construction of index codes using complex lattices}

\renewcommand{\arraystretch}{1.25}
\begin{table*}
\begin{minipage}[t]{0.5\linewidth}
\centering
\caption{All non-associate Gaussian primes of norm up to $53$}
\begin{tabular} {||c|c|c||}
\hline
Norm  & Prime & Rate \\
$|\phi|^2$ & $\phi$ & $\log_2 |\phi|$ \\
\hline
$2$     &  $1+i$ & $0.5$ \\
\hline
$5$ & $1+2i,1-2i$ & $1.16$ \\
\hline
$9$ & $3$ & $1.59$ \\
\hline
$13$ & $2+3i,2-3i$ & $1.85$ \\
\hline
$17$ & $1+4i,1-4i$ & $2.04$ \\
\hline
$29$ & $2+5i,2-5i$ & $2.43$ \\
\hline
$37$ & $1+6i,1-6i$ & $2.60$ \\
\hline
$41$ & $4+5i,4-5i$ & $2.68$ \\
\hline
$49$ & $7$ & $2.81$ \\
\hline
$53$ & $2+7i,2-7i$ & $2.86$ \\
\hline
\end{tabular}
\label{tb:Gaussian_primes}
\end{minipage}%
\begin{minipage}[t]{0.5\linewidth}
\centering
\caption{All non-associate Eisenstein primes of norm up to $61$}
\begin{tabular} {||c|c|c||}
\hline
Norm  & Prime & Rate \\
$|\phi|^2$ & $\phi$ & $\log_2|\phi|$ \\
\hline
$3$     &  $1-\om$ & $0.79$\\
\hline
$4$ & $2$ & $1$ \\
\hline
$7$ & $1+3\om,1+3\omb$ & $1.40$\\
\hline
$13$ & $1+4\om,1+4\omb$ & $1.85$\\
\hline
$19$ & $2+5\om,2+5\omb$ & $2.12$ \\
\hline
$25$ & $5$ & $2.32$ \\
\hline
$31$ & $1+6\om,1+6\omb$ & $2.48$ \\
\hline
$37$ & $3+7\om,3+7\omb$ & $2.60$ \\
\hline
$43$ & $1+7\om,1+7\omb$ & $2.71$ \\
\hline
$61$ & $4+9\om,4+9\omb$ & $2.97$ \\
\hline
\end{tabular}
\label{tb:Eisenstein_primes}
\end{minipage}
\end{table*}
\renewcommand{\arraystretch}{1}

Let $\Db$ be $\Zb[i]$ or $\Zb[\omega]$, and $\phi_1,\dots,\phi_K$ be any $K$ distinct non-associate primes in $\Db$. Let
\begin{align*}
M=\prod_{k=1}^{K}\phi_k, \text{ and } M_k=\sfrac{M}{\phi_k}=\prod_{\ell \neq k}\phi_{\ell} \text{ for } k=1,\dots,K.
\end{align*}
Let $\Lt$ be any $\Db$-lattice of real dimension $n$, and $\La=\Psi(\Lt)$ be its real version.
We construct our lattice index code by setting
\begin{align} \label{eq:complex_lattice_index_code}
\Ls=\Psi(M\Lt), \text{ and } \La_k=\Psi(M_k\Lt), \, k=1,\dots,K.
\end{align}
Since \mbox{$M_k \divides M$}, we have \mbox{$M\Lt \subset M_k\Lt$}, and hence, the coarse lattice $\Ls$ is a sub-lattice of each $\La_k$, \mbox{$k=1,\dots,K$}.
Using~\eqref{eq:vol_complex_lattice}, the message size of the $k^{\text{th}}$ symbol is
\begin{align*} 
 |\La_k/\Ls| = \frac{\vol(M\Lt)}{\vol(M_k\Lt)} = \frac{|M|^n\vol(\La)}{|M_k|^n\vol(\La)} = |\phi_k|^n,
\end{align*}
and its rate is
\begin{align*}
R_k = \sfrac{1}{n} \log_2 \left( |\phi_k|^n \right) = \log_2 |\phi_k| \text{ b/dim}.
\end{align*}

Tables~\ref{tb:Gaussian_primes} and~\ref{tb:Eisenstein_primes} list the first few non-associate Gaussian and Eisenstein primes, respectively. These are unique up to unit multiplication.
In Table~\ref{tb:Eisenstein_primes}, \mbox{$\omb=-1-\om$} is the complex conjugate of $\om=\exp\left(\sfrac{i2\pi}{3}\right)$.
The tables also show the norm $|\phi|^2$ of the prime $\phi$, and the corresponding message rate $\log_2|\phi|$ in~b/dim.

\begin{example}
The lattice $\La=D_4$ is a Gaussian lattice with the complex generator matrix
\begin{align*}
\widetilde{G} = \begin{pmatrix} 1 & 0 \\ 1 & 1+i\end{pmatrix}.
\end{align*}
Using~\eqref{eq:generator_Gaussian}, we obtain the $4 \times 4$ real generator matrix
\begin{align*}
G = \begin{pmatrix}
    1 & 0 & 0 & 0 \\
    1 & 1 & 0 & -1 \\
    0 & 0 & 1 & 0 \\
    0 & 1 & 1 & 1
    \end{pmatrix}.
\end{align*}
Let \mbox{$K=2$}, \mbox{$\phi_1=1+i$} and \mbox{$\phi_2=1+2i$}.
Then \mbox{$M=-1+3i$}, \mbox{$M_1=1+2i$} and \mbox{$M_2=1+i$}.
The real generator matrix of $\Psi(m\Lt)$ is \mbox{$\mathcal{M}(m) \times G$}.
The generator matrices of \mbox{$\La_1=\Psi(M_1\Lt)$}, \mbox{$\La_2=\Psi(M_2\Lt)$} and \mbox{$\Ls=\Psi(M\Lt)$}, thus obtained, are
\begin{align*}
G_1 &= \begin{pmatrix}[r] 1 & 0 & -2 & 0 \\ 1 & -1 & -2 & 3 \\ 2 & 0 & 1 & 0 \\ 2 & 3 & 1 & 1 \end{pmatrix}, \,
G_2 = \begin{pmatrix}[r] 1 & 0 & -1 & 0 \\ 1 & 0 & -1 & 2 \\ 1 & 0 & 1 & 0 \\ 1 & 2 & 1 & 0 \end{pmatrix} \text{ and } \\[5pt]
&~~~~~~~~~~~~~G_{\rm c} = \begin{pmatrix}[r] -1 & 0 & 3 & 0 \\ -1 & -4 & 3 & 2 \\ 3 & 0 & 1 & 0 \\ 3 & 2 & 1 & 4 \end{pmatrix},
\end{align*}
respectively. The message sizes are \mbox{$|\La_1/\Ls|=4$}, \mbox{$|\La_2/\Ls|=25$}, and the rates are \mbox{$R_1=\log_2|1+i|=\sfrac{1}{2}$~b/dim} and \mbox{$R_2=\log_2|1+2i|=\sfrac{1}{2}\log_2 5$~b/dim}. \hfill\IEEEQED
\end{example}

The following lemma will be useful in deriving the side information gain of the proposed lattice index codes.

\begin{lemma} \label{lem:gcd_M_S}
For every index set $S$, we have $\gcd(M_k,k \in S^{\mathsf{c}}) = \prod_{\ell \in S}\phi_{\ell}$.
\end{lemma}
\begin{IEEEproof}
Let \mbox{$d=\gcd(M_k,k \in S^{\mathsf{c}})$}.
Since each $M_k$ is a product of a subset of the primes $\phi_1,\dots,\phi_K$, \mbox{$d=\gcd(M_k,k \in S^{\mathsf{c}})$} is of the form $\phi_1^{e_1}\cdots\phi_K^{e_K}$ with \mbox{$e_k \in \{0,1\}$}. If \mbox{$k \in S^{\mathsf{c}}$}, we have \mbox{$d\divides M_k$}, and since $\phi_k$ is not a factor of $M_k$, we obtain \mbox{$e_k=0$}. It follows that $d\divides \prod_{\ell \in S} \phi_{\ell}$. On the other hand, it is easy to verify that $\prod_{\ell \in S} \phi_{\ell}\divides M_k$ for every \mbox{$k \in S^{\mathsf{c}}$}, implying that $\prod_{\ell \in S}\phi_{\ell}\divides d$. Hence, \mbox{$d=\prod_{\ell \in S}\phi_{\ell}$}.
\end{IEEEproof}

We now show, in Lemma~\ref{lem:arbit_construction}, that the lattice index code $\mathscr{C}$ is $\La/\Ls$ and the encoding map $\rho$ is injective.
Part~\emph{(\ref{lem:part3:arbit_construction})} of Lemma~\ref{lem:arbit_construction} will later allow us to show that the minimum distance $d_S$ with side information index set $S$ is exponential in $R_S$.

\begin{lemma} \label{lem:arbit_construction}
With the lattices $\La_1,\dots,\La_K$ and $\Ls$ defined as~\eqref{eq:complex_lattice_index_code},
\begin{enumerate}[(i)]
\item the encoding map $\rho$ in Definition~\ref{def:lattice_index_code} generates a lattice index code with transmit codebook $\mathscr{C}=\La/\Ls$; and \label{lem:part1:arbit_construction}
\item for any $S$, we have \mbox{$\sum_{k \in S^{\mathsf{c}}}\La_k = \Psi\left( \prod_{\ell \in S}\phi_{\ell} \, \Lt\right)$}. \label{lem:part3:arbit_construction}
\end{enumerate}
\end{lemma}
\begin{IEEEproof}
See Appendix~\ref{app:lem:arbit_construction}
\end{IEEEproof}


\begin{figure*}[!t]
\centering
\includegraphics[width=4in]{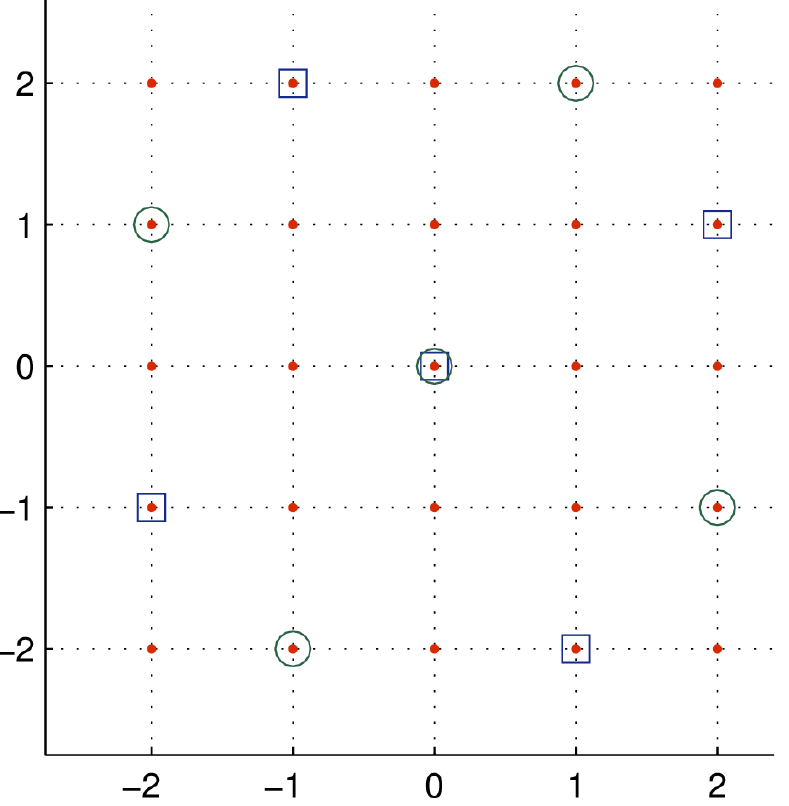}
\caption{The constellation of Example~\ref{ex:25QAM}. The dots constitute the code \mbox{$\mathscr{C}=25\text{-QAM}$}, the squares and circles correspond to $\La_1/\Ls$ and $\La_2/\Ls$, respectively.}
\label{fig:25QAM}
\end{figure*}


\begin{lemma} \label{lem:dS_RS}
For every choice of $S$, \mbox{$R_S=\log_2 \left( \sfrac{d_S}{d_0} \right)$}, and hence the side information gain is uniform.
\end{lemma}
\begin{IEEEproof}
Using~\eqref{eq:dS_lambda},~\eqref{eq:dmin_complex_lattice} and Part~\emph{(\ref{lem:part3:arbit_construction})} of Lemma~\ref{lem:arbit_construction}, we have
\begin{align}
d_S = d_{\min} \left(\sum_{k \in S^{\mathsf{c}}} \La_k \right) &= d_{\min} \left( \Psi \left( \prod_{\ell \in S} \phi_{\ell} \Lt \right) \right) \nonumber \\ &= \prod_{\ell \in S} \left| \phi_{\ell} \right| \, d_0. \label{eq:dS}
\end{align}
The side information rate corresponding to $S$ is
\begin{align} \label{eq:RS}
 R_S = \sum_{k \in S} R_k = \sum_{k \in S} \log_2 |\phi_k| = \log_2 \left( \prod_{k \in S} |\phi_k| \right).
\end{align}
From~\eqref{eq:dS} and~\eqref{eq:RS}, we see that \mbox{$R_S = \log_2 \left(\sfrac{d_S}{d_0}\right)$} for every choice of $S$, and $10 \log_{10}\left(\sfrac{d_S^2}{d_0^2}\right)/R_S$ is independent of $S$.
\end{IEEEproof}

Using the relation \mbox{$R_S=\log_2 \left( \sfrac{d_S}{d_0} \right)$} with~\eqref{eq:side_inf_gain}, we obtain \mbox{$\Gamma(\mathscr{C}) \approx 6$~dB/b/dim}. Thus, when $\La$ is the densest lattice in its dimension, the proposed construction achieves the optimal side information gain over all lattice index codes constructed based on $\La$.
Note that this optimality with respect to $\Gamma$ holds only among the family of lattice index codes of Definition~\ref{def:lattice_index_code}, and when $\La$ is densest in its dimension.
While Example~\ref{ex:lattice_more_than_6} gives a lattice index code with \mbox{$\Gamma > 6$~dB/b/dim} using a lattice $\La$ that does not have highest density, Example~\ref{ex:16PSK} shows an index code with \mbox{$\Gamma > 6$~dB/b/dim} using a \emph{non}-lattice constellation.

\begin{example}[\emph{A $2$-message constellation using $25$-QAM}] \label{ex:25QAM}
Consider the non-associate primes \mbox{$\phi_1=1+2i$} and \mbox{$\phi_2=1-2i$} in \mbox{$\Db=\Zb[i]$}.
Setting $$\Lt=\Zb[i],$$ we obtain a constellation $\mathscr{C}$ carved from $\La=\Psi(\Zb[i])=\Zb^2$.
We have \mbox{$M=\phi_1\phi_2=5$}, \mbox{$M_1=1-2i$} and \mbox{$M_2=1+2i$}.
The coarse lattice \mbox{$\Psi(5\Zb[i])=5\Zb^2$}, and the lattice index code
\begin{align*}
\mathscr{C} = \Psi\left(\Zb[i]\right)/\Psi\left(5\Zb[i]\right) = \Zb^2/5 \Zb^2
\end{align*}
is the $25$-QAM constellation.
The generator matrices of the lattices $\La_1=\Psi(M_1\Zb[i])$ and $\La_2=\Psi(M_2\Zb[i])$ are
\begin{align*}
\begin{pmatrix}[r] 1 & 2 \\ -2 & 1 \end{pmatrix} \text{ and } \begin{pmatrix}[r] 1 & -2 \\ 2 & 1 \end{pmatrix},
\end{align*}
respectively. The constellations \mbox{$\La_1/\Ls$} and \mbox{$\La_2/\Ls$} consist of $5$ points each (see Fig.~\ref{fig:25QAM}),
\begin{align*}
 \La_1/\Ls &= \left\{ 0, (1,-2)^\intercal,(2,1)^\intercal,(-2,-1)^\intercal,(-1,2)^\intercal  \right\}, \\
 \La_2/\Ls &= \left\{ 0,(1,2)^\intercal,(2,-1)^\intercal,(-2,1)^\intercal,(-1,-2)^\intercal \right\}.
\end{align*}
The minimum squared distance of $\La$ is $1$, while that of $\La_1$ and $\La_2$ is $5$. When the side information index set is \mbox{$S=\{1\}$} or $\{2\}$, the squared distance gain is $10\log_{10}5$~dB, and the side information rate $R_S=\sfrac{1}{2}\log_2 5$~b/dim, yielding a side information gain of \mbox{$\Gamma \approx 6$~dB/b/dim}.
Fig.~\ref{fig:25QAM_simulation} shows the performance of the three different receivers with \mbox{$S=\varnothing$} (no side information), \mbox{$S=\{1\}$}, and \mbox{$S=\{2\}$}, respectively. The performance for \mbox{$S=\{1\}$} and \mbox{$S=\{2\}$} were obtained by simulations, while that for \mbox{$S=\varnothing$} was obtained through the closed form expression for the error rate of $25$-QAM~\cite{Proakis_01}.
From the simulation result, we observe that at the error rate of $10^{-5}$, the knowledge of either of the two transmitted messages provides an ${\sf SNR}$ gain of $6.95$~dB. When normalized by the side information rate $\sfrac{1}{2}\log_2 5$~b/dim, we have a normalized ${\sf SNR}$ gain of $5.98$~dB/b/dim, which is a good match with $\Gamma \approx 6$~dB/b/dim. \hfill\IEEEQED
\begin{figure*}[!t]
\centering
\includegraphics[width=5in]{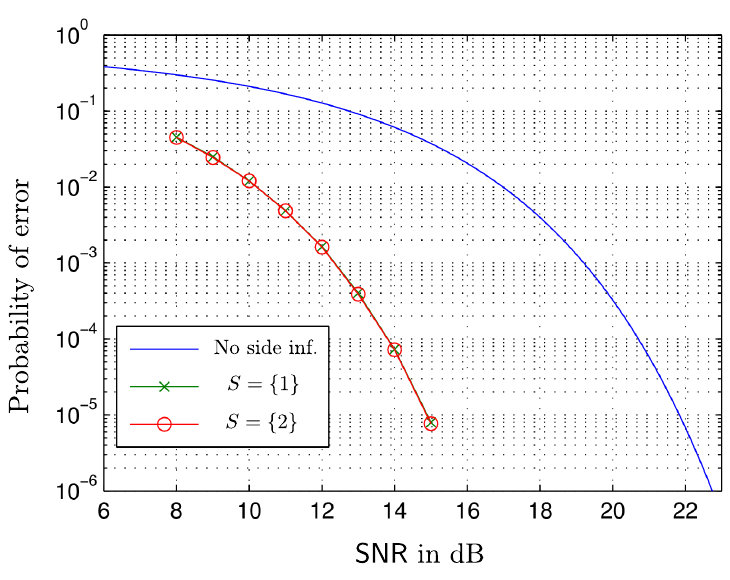}
\caption{Performance of the code of Example~\ref{ex:25QAM} for three different receivers.}
\label{fig:25QAM_simulation}
\end{figure*}
\end{example}

\subsection*{Construction with $\Db=\Zb$}

Let \mbox{$p_1,\dots,p_K \in \Zb$} be distinct rational primes, \mbox{$M=p_1 \cdots p_K$} be their product and \mbox{$M_k=\sfrac{M}{p_k}$}, \mbox{$k=1,\dots,K$}. Let \mbox{$\La \subset \Rb^n$} be any $n$-dimensional lattice. We let
\begin{align*}
\Ls = M \La \text{ and } \La_k = M_k \La.
\end{align*}
The rate of the $k^{\text{th}}$ message is
\begin{align*}
 R_k = \frac{1}{n} \log_2 \left( \frac{\vol\left( M \La \right)}{\vol\left( M_k \La \right)} \right) = \frac{1}{n} \log_2 p_k^n = \log_2 p_k.
\end{align*}
Similar to Lemmas~\ref{lem:arbit_construction} and~\ref{lem:dS_RS}, we can show that $\mathscr{C}=\La/\Ls$, $\rho$ is injective, $R_S=\log_2\left(\sfrac{d_S}{d_0}\right)$, and hence, $\Gamma \approx 6$~dB/b/dim.

\begin{example}
The code of Example~\ref{ex:toy_example} can be obtained by using \mbox{$\Db=\Zb$}, \mbox{$\La=\Zb$}, and the tuple of prime numbers \mbox{$(\phi_1,\phi_2,\phi_3)=(2,3,5)$}. \hfill\IEEEQED
\end{example}

A construction of lattice codes using tuples of prime integers in $\Zb[i]$ and $\Zb[\omega]$ is reported in~\cite{HNT_arxiv_14} for low complexity multilevel encoding and multistage decoding in compute-and-forward applications.

When $\La$ is a Gaussian or Eisenstein lattice, the message rates available from the proposed lattice index codes are $\log_2|\phi|$~b/dim, where \mbox{$\phi \in \Db$} is prime~(see Tables~\ref{tb:Gaussian_primes} and~\ref{tb:Eisenstein_primes}). When $\Db=\Zb$, the codes allow one message of rate $\log_2 p$~b/dim for every rational prime \mbox{$p \in \Zb$}. In Section~\ref{sec:Hurwitz} we construct a family of lattice index codes from a class of quaternionic lattices, which includes $D_4^*$ and $E_8$, that allow encoding two messages, of rate $\sfrac{1}{2} \log_2 p$~b/dim each, for every odd rational prime $p \in \Zb$.
The codes of Section~\ref{sec:Hurwitz} thus provide further choices in terms of message rates at the source and side information rates at the receivers.

\section{Construction of lattice index codes using Hurwitz Integers} \label{sec:Hurwitz}

We construct lattice index codes using quaternionic lattices by exploiting the fact that the Hurwitz integral quaternions $\Hb$ form a non-commutative PID. Since the ideals in $\Hb$ are not two-sided in general, the Chinese remainder theorem does not apply to $\Hb$. However, we identify a set of ideals that lead to uniform gain lattice index codes with side information gain \mbox{$\sim 6$~dB/b/dim}.

We first consider the one dimensional $\Hb$-lattice $D_4^*$ in Section~\ref{sec:D4}, and then extend the results to a class of higher dimensional $\Hb$-lattices in Section~\ref{sec:quat_lattices}.
We now briefly review some properties of the Hurwitz integers $\Hb$. We refer the reader to~\cite{CoS_Peters_03} for more details.

\subsection{Review of Hurwitz integers}

The set of Hurwitz integers $\Hb$ is the subring of quaternions consisting of those elements whose coordinates are either all in $\Zb$ or all in $\Zb + \sfrac{1}{2}$, i.e.,
\begin{align*}
\Hb = &\left\{ a + bi + cj + dk~\big\vert~ a,b,c,d \in \Zb \right\} \\ &~~~~~~~~~~~~
\bigcup \left\{ a + bi + cj + dk~\big\vert~ a,b,c,d \in \Zb + \sfrac{1}{2} \right\}.
\end{align*}
Addition in $\Hb$ is component-wise, and multiplication is defined by the relations
\mbox{$i^2=j^2=-1$} and \mbox{$ij=-ji=k$}.
This makes $\Hb$ non-commutative. For \mbox{$A=a+bi+cj+dk \in \Hb$}, the conjugate of $A$ is {$\overline{A}=a-bi-cj-dk$}, and the norm is
\begin{align*}
N(A)=A\overline{A}=\overline{A}A=a^2+b^2+c^2+d^2 \in \Zb.
\end{align*}
The real part of $A$ is \mbox{$\real(A)=a$}, and the trace is \mbox{$A + \overline{A} = 2a$}.
The \emph{four-square theorem} of Lagrange states that every positive integer is a sum of four integer-squares, i.e., every positive integer is the norm of some Hurwitz integer.
The units of $\Hb$ are the elements with norm $1$. There are precisely $24$ units in $\Hb$, eight of them $\pm 1, \pm i, \pm j, \pm k$ have integer coordinates, and the remaining $16$ units $\pm \frac{1}{2} \pm \frac{i}{2} \pm \frac{j}{2} \pm \frac{k}{2}$ have half-integer coordinates.

The ring $\Hb$ is a Euclidean domain, and hence it is a non-commutative PID. Every left ideal $I$ of $\Hb$ is generated by a single element, and is of the form \mbox{$I=\Hb A$} for some \mbox{$A \in \Hb$}. Similarly every right ideal is of the form \mbox{$I=A\Hb$}. In the rest of this section we will use only the left ideals in $\Hb$ to construct our constellations. Similar results can be obtained from right ideals. The generator of a (left) ideal is unique up to left multiplication by a unit of $\Hb$.

When viewed as a $4$-dimensional lattice, in the basis $\{1,i,j,k\}$, $\Hb$ yields $D_4^*$, and its generator matrix is
\begin{align*}
G=\begin{pmatrix}
\sfrac{1}{2} & 0 & 0 & 0 \\[3pt]
\sfrac{1}{2} & 1 & 0 & 0 \\[3pt]
\sfrac{1}{2} & 0 & 1 & 0 \\[3pt]
\sfrac{1}{2} & 0 & 0 & 1
\end{pmatrix}.
\end{align*}
For \mbox{$A=a+bi+cj+dk$}, let \mbox{$\vect(A)=(a,b,c,d)^\intercal$} be the vector of the coordinates of $A$ in the basis $\{1,i,j,k\}$. For any \mbox{$B \in \Hb$}, we have \mbox{$\vect(BA) = \mathcal{M}(A) \vect(B)$}, where
\begin{align} \label{eq:matrix_of_quaternion}
\mathcal{M}(A) = \begin{pmatrix}[r] a & -b & -c & -d \\ b & a & d & -c \\ c & -d & a & b \\ d & c & -b & a  \end{pmatrix}.
\end{align}
Note that $\mathcal{M}(A)$ is an orthogonal matrix, and its determinant is \mbox{$(a^2+b^2+c^2+d^2)^2=N(A)^2$}.
The ideal \mbox{$\Hb A$} generated by $A$ is a sub-lattice of $D_4^*$, and its generator matrix is $\mathcal{M}(A)G$, where $G$ is the generator matrix of $D_4^*$, and $\mathcal{M}(A)$ corresponds to left multiplication of a quaternion by $A$. Thus, the volume of the fundamental region of the lattice $\Hb A$ is
\begin{align} \label{eq:H_volume}
\vol\left( \Hb A \right) = \left\vert \det\mathcal{M}(A) \right\vert \, \left\vert \det G \right\vert = \frac{N(A)^2}{2} .
\end{align}

The norm operation is multiplicative on $\Hb$, i.e., $N(AB)=N(A)N(B)$ for every \mbox{$A,B \in \Hb$}. The units of $\Hb$ are the elements with the shortest norm, and $N(A) \geq 1$ for $A \in \Hb$.
Let \mbox{$I=\Hb D$} be the ideal generated by the element $D$, and \mbox{$B \in I$}. Then, \mbox{$B=AD$} for some \mbox{$A \in \Hb$}, and its norm satisfies $$N(B)=N(AD)=N(A)N(D) \geq N(D).$$
Hence, the generator of $I$ is a shortest vector in the lattice $I$, and the minimum squared distance between any two points in \mbox{$I=\Hb D$} equals the norm $N(D)$ of the generator.

For \mbox{$A,B \in \Hb$}, we say that $A\divides B$ if \mbox{$B \in \Hb A$}, i.e., if $B$ belongs to the ideal generated by $A$.
If $A\divides B$, we have \mbox{$B=DA$} for some \mbox{$D \in \Hb$} and hence $N(A)\divides N(B)$.
The $\gcd$ of two elements $A$ and $B$ is the generator of the ideal generated by $A$ and $B$, i.e., \mbox{$\Hb A + \Hb B = \Hb \gcd(A,B)$}.
If \mbox{$D=\gcd(A,B)$}, we have $N(D)\divides N(A)$ and $N(D)\divides N(B)$ in $\Zb$, hence $N(D)\divides \gcd(N(A),N(B))$ in $\Zb$.

\subsection{Construction of lattice index codes based on $D_4^*$} \label{sec:D4}

Consider $L$ distinct \emph{odd} rational primes \mbox{$p_1,\dots,p_L \in \Zb$}.
From the four-square theorem~\cite{CoS_Peters_03}, there exist \mbox{$P_1,\dots,P_L \in \Hb$} such that \mbox{$p_i = N(P_i)$}. In order to prove the injectivity of $\rho$, we  further require that the real parts of the $P_i$'s be powers of $2$~(this technical assumption is used in the proof of Lemma~\ref{lem:H1}).
Using Legendre's \emph{three-square theorem}~\cite{UsH_McGraw_39}, we prove in Appendix~\ref{app:hurwitz_prime_real_part} that for every odd rational prime $p$ there exists a Hurwitz integer $P$ such that \mbox{$p=N(P)$} and $\real(P)$ is a power of $2$.
In particular, the proof only requires that $p$ be a positive odd rational integer (not necessarily a prime), and shows that $P$ can be chosen such that \mbox{$\real(P) \in \{1,2\}$}.

Define \mbox{$K=2L$} elements $M_1,\dots,M_K$, as 
\begin{align*}
M_k &= P_k \prod_{\ell \neq k} p_{\ell}, \text{ and } 
M_{k+L} = \overline{M}_k = \overline{P}_k \prod_{\ell \neq k} p_{\ell},
\end{align*}
for \mbox{$k=1,\dots,L$}. Let {$M=p_1 \cdots p_L$} be the generator of the ideal \mbox{$\Is=\Hb M$}.
Note that for each {$k=1,\dots,L$}, we have $M_k\divides M$ and $M_{k+L}\divides M$ since
\begin{align*}
M=p_1 \cdots p_L &= p_k  \prod_{\ell \neq k}p_{\ell}  = \overline{P}_k P_k  \prod_{\ell \neq k}p_{\ell} = P_k \overline{P}_k  \prod_{\ell \neq k}p_{\ell}, \\
\text{i.e., } M &= \overline{P}_k M_k = P_k M_{k+L}.
\end{align*}
Hence, $\Is=\Hb M$ is a sub-ideal of $\Hb M_k$, \mbox{$k=1,\dots,K$}. 
We use \mbox{$\Ls=\Is$}, and \mbox{$\La_k=\Hb M_k$}, \mbox{$k=1,\dots,K$}, in Definition~\ref{def:lattice_index_code} to construct our lattice index code.
Using~\eqref{eq:H_volume}, 
\begin{align} \label{eq:size_Wk}
|\Hb M_k /\Hb M| = \frac{\vol\left( \Hb M \right)}{\vol\left( \Hb M_k \right)} = \frac{N(M)^2}{N(M_k)^2} = \begin{cases} p_k^2, &  k \leq L, \\ p_{k-L}^2, &  k > L.  \end{cases}
\end{align}
Since $\Hb$ is a $4$-dimensional lattice, the rate of the $k^{\text{th}}$ message is
\begin{align*}
R_k = \frac{\log_2 |\Hb M_k/\Hb M|}{4}=\begin{cases} \sfrac{1}{2} \log_2 p_k, &  k \leq L, \\ \sfrac{1}{2} \log_2 p_{k-L}, &  k > L.  \end{cases}
\end{align*}
The side information rate for \mbox{$S \subset \{1,\dots,K\}$} is
\begin{align*}
R_S = \sum_{k \in S} R_k = \frac{1}{4} \log_2 \left( \prod_{k \in S} |\Hb M_k/\Hb M|  \right) \text{ b/dim}.
\end{align*}
Table~\ref{tb:Hurwitz_primes} provides one instance (among many possible) of Hurwitz integer $P$ with {$N(P)=p$} and {$\real(P)=2^m$} for each of the first ten odd primes $p$.
Table~\ref{tb:Hurwitz_primes} also lists the message rate $\sfrac{1}{2} \log_2 p$~b/dim available from using each Hurwitz integer $P$.

\renewcommand{\arraystretch}{1.25}
\begin{table}
\centering
\caption{Examples of Hurwitz integers with odd-prime norm and real part a power of $2$}
\begin{tabular} {||c|c|c||}
\hline
Norm  & Hurwitz integer & Rate \\
$N(P)=p$ & $P$ & $\sfrac{1}{2} \log_2 p$ \\
\hline
$3$     &  $1+i+j$ & $0.79$ \\
\hline
$5$ & $1+2i$ & $1.16$ \\
\hline
$7$ & $1+i+j+2k$ & $1.40$ \\
\hline
$11$ & $1+i+3j$ & $1.73$ \\
\hline
$13$ & $2+3i$ & $1.85$ \\
\hline
$17$ & $1+4i$ & $2.04$ \\
\hline
$19$ & $1+3i+3j$ & $2.12$ \\
\hline
$23$ & $1+2i+3j+3k$ & $2.26$ \\
\hline
$29$ & $2+5i$ & $2.43$ \\
\hline
$31$ & $1+i+2j+5k$ & $2.48$ \\
\hline
\end{tabular}
\label{tb:Hurwitz_primes}
\end{table}
\renewcommand{\arraystretch}{1}

\begin{example} \label{ex:Hurwitz}
Consider $L=2$ and the odd primes \mbox{$p_1=3$} and \mbox{$p_2=5$}. With \mbox{$P_1=1+i+j$} and \mbox{$P_2=1+2i$}, we have \mbox{$p_k=N(P_k)$} and \mbox{$\real(P_k)=1=2^0$}. We have \mbox{$K=2L=4$} information symbols with constellations \mbox{$\Hb M_k/ \Hb M$}, where \mbox{$M=p_1p_2=15$},
\begin{align*}
 M_1 &= P_1p_2 = 5(1+i+j),  \,  M_2 = P_2p_1 = 3(1+2i) \\
 M_3 &= \overline{M}_1 = 5(1-i-j), \text{ and }  M_4 = \overline{M}_2 = 3(1-2i).
\end{align*}
The cardinalities of the four constellations are $9,25,9$ and $25$, respectively, and their rates are $\sfrac{1}{2}\log_2 3$, $\sfrac{1}{2}\log_2 5$, $\sfrac{1}{2}\log_2 3$, and $\sfrac{1}{2}\log_2 5$~b/dim. \hfill\IEEEQED
\end{example}

In the rest of this sub-section we show that the choice
\begin{align*}
\Ls = \Is = \Hb M \text{ and } \La_k = \Hb M_k, \, k=1,\dots,K,
\end{align*}
produces a uniform gain lattice index code with side information gain \mbox{$\sim 6$~dB/b/dim}. We show that the transmit codebook $\mathscr{C}$ equals $\Hb/\Is$~(Lemma~\ref{lem:H1}), the encoding map $\rho$ is injective~(Lemma~\ref{lem:rho_bijective_H}), and the minimum distance $d_S$ is exponential in the side information rate $R_S$~(Lemma~\ref{lem:dS_RS_H}).

\begin{lemma} \label{lem:H1}
The transmit codebook $\mathscr{C}$ equals $\Hb/\Is$.
\end{lemma}
\begin{IEEEproof}
See Appendix~\ref{app:lem:H1}
\end{IEEEproof}

\begin{lemma} \label{lem:rho_bijective_H}
The map $\rho: \Hb M_1/\Is \times \cdots \times \Hb M_K/\Is \to \mathscr{C}$ is injective.
\end{lemma}
\begin{IEEEproof}
It is enough to show that \mbox{$|\Hb M_1/\Is \times \cdots \times \Hb M_K/\Is|=|\Hb/\Is|$}. From~\eqref{eq:size_Wk},
\begin{align} \label{eq:cardinality_norm_M_H}
|\Hb M_1/\Is \times \cdots \times \Hb M_K/\Is|=\left(\prod_{k=1}^{L}p_k^2\right)^2=N(M)^2.
\end{align}
Also, $$|\Hb/\Is| = \sfrac{\vol\left( \Hb M \right)}{\vol\left( \Hb \right)} = N(M)^2.$$
\end{IEEEproof}

The minimum squared distance $d_S^2$ corresponding to $S$ satisfies \mbox{$d_S^2 = d_{\min}^2\left( \sum_{k \in S^{\mathsf{c}}} \Hb M_k \right)$}.
Denoting the generator of the ideal $\sum_{k \in S^{\mathsf{c}}}\Hb M_k$ by $D_S$, we have \mbox{$d_S^2=N(D_S)$}.

\begin{lemma} \label{lem:dS_RS_H}
For every choice of $S$, we have \mbox{$R_S=\log_2 d_S$}, and hence the side information gain is uniform.
\end{lemma}
\begin{IEEEproof}
Consider the restriction $\rho \vert_{S^\mathsf{c}}$ of the encoding map $\rho$, in~\eqref{eq:rho}, to the subset of messages with indices in $S^\mathsf{c}$, i.e.,
\begin{align*}
 \rho \vert_{S^\mathsf{c}} \left( x_k, k \in S^{\mathsf{c}} \right) = \sum_{k \in S^{\mathsf{c}}} x_k {\rm~mod~} \Is.
\end{align*}
The image of $\rho\vert_{S^\mathsf{c}}$ is {$\sum_{k \in S^{\mathsf{c}}}\Hb M_k/\Is=\Hb D_S/\Is$}, where $D_S$ is the generator of the ideal $\sum_{k \in S^{\mathsf{c}}}\Hb M_k$.
Since $\rho$ is injective (Lemma~\ref{lem:rho_bijective_H}), so is its restriction $\rho \vert_{S^\mathsf{c}}$. Hence, the domain and the image of $\rho\vert_{S^\mathsf{c}}$ have the same cardinality, i.e.,
\begin{align*}
\prod_{k \in S^{\mathsf{c}}} |\Hb M_k/\Is| = |\Hb D_S / \Is| = \frac{N(M)^2}{N(D_S)^2}
\end{align*}
Using~\eqref{eq:cardinality_norm_M_H} with the above equation, we get
\begin{align} \label{eq:norm_DS}
N(D_S)^2 = \prod_{k \in S} |\Hb M_k/\Is| = \prod_{k \in S} 2^{4R_k} = 2^{4R_S}.
\end{align}
Substituting $N(D_S)=d_S^2$ we obtain the desired result.
\end{IEEEproof}

Using Lemma~\ref{lem:dS_RS_H} and \mbox{$d_0=d_{\min}(\Hb)=1$} in~\eqref{eq:side_inf_gain} we see that the side information gain of the proposed constellation equals the upper bound \mbox{$\sim 6$~dB/b/dim}, and it satisfies the uniform gain condition~\eqref{eq:uniform_gain}.

\subsection{Construction of index codes using quaternionic lattices} \label{sec:quat_lattices}

We first recall the definition of \emph{quaternionic lattices}, and then show that the extension of the technique used in Section~\ref{sec:D4} to those quaternionic lattices which are two-sided $\Hb$-modules produces uniform gain lattice index codes.

\subsubsection*{Quaternionic lattices}

We denote the quaternion algebra by
$$\Quat=\left\{a+bi+cj+dk \, \vert \, a,b,c,d \in \Rb\right\}.$$
A quaternionic lattice $\Lt$ of dimension $t$ over $\Quat$ is a discrete left-$\Hb$ sub-module of $\Quat^t$~\cite{CoS_Springer_99}, i.e., $A \Lt \subset \Lt$ for every $A \in \Hb$, where
\begin{align*}
A \Lt = \left\{ \left(AV_1,\dots,AV_t\right)^\intercal \, \big\vert \, \left(V_1,\dots,V_t\right)^\intercal \in \Lt \right\rbrace.
\end{align*}
The real lattice $\La$ associated with $\Lt$ is obtained by the map \mbox{$\Psi:\Quat^t \to \Rb^{4t}$}, where $\Psi\left((V_1,\dots,V_t)^\intercal\right)$ is the real vector consisting of the $\{1,i,j,k\}$-coordinates of each of the $t$ quaternions $V_1,\dots,V_t$. Hence, the real dimension of $\Lt$ is \mbox{$n=4t$}. Note that \mbox{$\Psi(\Lt_1) \subset \Psi(\Lt_2)$} if and only if \mbox{$\Lt_1 \subset \Lt_2$}, and \mbox{$\Psi(\Lt_1+\Lt_2)=\Psi(\Lt_1)+\Psi(\Lt_2)$}.

\begin{example}
The Gosset lattice $E_8$ is the real version of a quaternionic lattice $\Lt$ of dimension \mbox{$t=2$} over $\Hb$~\cite{CoS_Springer_99}. Its generator matrix over $\Hb$ is
\begin{align*}
\begin{pmatrix} 1+i & 1 \\ 0 & 1 \end{pmatrix}.
\end{align*}
The lattice $\Lt \subset \Quat^2$ consists of all left $\Hb$-linear combinations of the two columns of this generator matrix, i.e.,
\begin{align} \label{eq:E8}
\Lt = \left\{ \begin{pmatrix} A(1+i) + B \\ B \end{pmatrix} \, \Big\vert \, A,B \in \Hb  \right\}.
\end{align}

\hfill\IEEEQED
\end{example}

Some of the well known high-density lattices, such as $D_4^*$, $D_4$, $E_8$, $\La_{12}^{\max}$ and $\La_{24}$ can be viewed as quaternionic lattices~\cite{CoS_Springer_99}. The lattice index codes of Section~\ref{sec:D4} were built using the one-dimensional quaternionic lattice $D_4^*$. A direct extension of this construction to arbitrary higher dimensional quaternionic lattices, as conducted in Section~\ref{sec:commutative} for complex lattices, does not appear to hold because of the non-commutativity of $\Hb$. The problem arises in determining if one lattice is a subset of another.
Given a $\Hb$-lattice $\Lt$, we construct the component lattices of our index code by right-multiplying $\Lt$ with appropriate Hurwitz integers. Consider
\begin{align*}
\Lt M = \left\{ \left(V_1 M,\dots,V_t M\right)^\intercal \, \big\vert \, \left(V_1,\dots,V_t\right)^\intercal \in \Lt \right\rbrace,
\end{align*}
where $M \in \Hb$. Since $M$ multiplies on the right, $\Lt M$ inherits the property of being a left-$\Hb$ module from $\Lt$, and hence, it is a quaternionic lattice.
In our construction, for any $M_k,M \in \Hb$ with $M_k \divides M$, we require that \mbox{$\Lt M \subset \Lt M_k$}. If \mbox{$M=AM_k$}, this condition translates to \mbox{$\Lt A M_k \subset \Lt M_k$}, which can be guaranteed if \mbox{$\Lt A \subset \Lt$}, i.e., if $\Lt$ is a right-$\Hb$ module in addition to being a left-$\Hb$ module. In the rest of this section we assume that $\Lt$ is a two-sided $\Hb$ module.
As an example, we now show that $E_8$ is a two-sided $\Hb$-module, and hence can be used as the base lattice $\Lt$ in our construction.

\begin{lemma}
The Gosset lattice $E_8$ is a right-$\Hb$ module.
\end{lemma}
\begin{IEEEproof}
Let $\Lt$, as defined in~\eqref{eq:E8}, be the quaternionic version of $E_8$. Consider
\begin{align*}
\Lt_{\sf right} = \left\{ \begin{pmatrix} (1+i)C + D \\ D \end{pmatrix} \, \Big\vert \, C,D \in \Hb  \right\}.
\end{align*}
It is clear that $\Lt_{\sf right}$ is a right-$\Hb$ module. We will complete the proof by showing that \mbox{$\Lt=\Lt_{\sf right}$}. In order to prove the equality of the two sets, we need to show that for every \mbox{$A,B \in \Hb$} there exist \mbox{$C,D \in \Hb$} such that $$\left(\, A(1+i)+B,B \,\right)^\intercal=\left(\, (1+i)C+D,D \,\right)^\intercal,$$ and vice versa. This is valid if and only if \mbox{$B=D$} and \mbox{$A(1+i)=(1+i)C$}. If \mbox{$A=a+bi+cj+dk$}, a direct computation shows that \mbox{$C=a+bi+dj-ck$} satisfies \mbox{$A(1+i)=(1+i)C$}. This completes the proof.
\end{IEEEproof}

Right multiplying each component of $V=(V_1,\dots,V_t) \in \Lt$ by $M$ is equivalent to left multiplying the real vector $\Psi(V)$ by the $4t \times 4t$ matrix
\begin{align} \label{eq:matrix_mult_quaternion_lattice}
\begin{pmatrix}[c]
\mathcal{M}(M) & & & \\
  & \mathcal{M}(M) &  &  \\
  &     &  \ddots &  \\
  &     &         & \mathcal{M}(M)
\end{pmatrix},
\end{align}
which consists of $t$ copies of the matrix $\mathcal{M}(M)$, and where the function $\mathcal{M}(\cdot)$ is given in~\eqref{eq:matrix_of_quaternion}.
The generator matrix of $\Psi(\Lt M)$ is the product of~\eqref{eq:matrix_mult_quaternion_lattice} and the generator matrix of $\Psi(\Lt)$.
Since $\mathcal{M}(M)$ is orthogonal with determinant $N(M)^2$, the matrix~\eqref{eq:matrix_mult_quaternion_lattice} is orthogonal with determinant $N(M)^{2t}$. Hence, the volume and the squared minimum distance of the lattice $\Psi(\Lt M)$ are
\begin{align*}
\vol(\Lt M) &= \vol\left( \Psi(\Lt M) \right) = N(M)^{2t} \, \vol\left(\Psi(\Lt)\right), \\
d_{\min}^2(\Lt M) &= d_{\min}^2\left( \Psi(\Lt M) \right) = N(M) \, d_{\min}^2\left( \Psi(\Lt) \right).
\end{align*}

\subsubsection*{Construction on two-sided $\Hb$-modules}

The following lemma enables us to extend the construction of Section~\ref{sec:D4} to all lattices $\Lt$ that are two-sided $\Hb$-modules.

\begin{lemma} \label{lem:quat_enabling_lemma}
If $A,B \in \Hb$ are such that $A \divides B$, then $\Lt A \supset \Lt B$.
\end{lemma}
\begin{IEEEproof}
Let \mbox{$B=DA$} and \mbox{$\lambda \in \Lt B$}. Then \mbox{$\lambda=VB$} for some \mbox{$V \in \Lt$}, and hence, \mbox{$\lambda=VB=VDA$}. Since $\Lt$ is a right-$\Hb$ module, \mbox{$VD \in \Lt$}, and hence \mbox{$\lambda \in \Lt A$}.
\end{IEEEproof}

Let $M_1,\dots,M_K$ and $M$ be as defined in Section~\ref{sec:D4}.
We set
\begin{align*}
\Lt_k=\Lt M_k,~k \in 1,\dots,K, \text{ and } \Lts=\Lt M.
\end{align*}
We construct our quaternionic lattice index code by using
\begin{align*}
\Ls = \Psi\left(\Lts\right) = \Psi\left(\Lt M\right) \text{ and }
\La_k = \Psi\left(\Lt_k\right) = \Psi\left(\Lt M_k\right).
\end{align*}
Since $M_k \divides M$, using Lemma~\ref{lem:quat_enabling_lemma}, we have $\Lts \subset \Lt_k$, and hence $\Ls \subset \La_k$, for all $k=1,\dots,K$.
The cardinality $|\La_k/\Ls|$ of the $k^{\text{th}}$ message is
\begin{align*}
\frac{\vol(\Ls)}{\vol(\La_k)} = \frac{\vol(\Lt M)}{\vol(\Lt M_k)} = \frac{N(M)^{2t}}{N(M_k)^{2t}} = \begin{cases} p_k^{2t}, &  k \leq L, \\ p_{k-L}^{2t}, &  k > L,  \end{cases}
\end{align*}
and the rate is
\begin{align*}
R_k = \frac{1}{4t} \log_2 |\La_k/\Ls|  = \begin{cases} \sfrac{1}{2} \log_2 p_k, &  k \leq L, \\ \sfrac{1}{2} \log_2 p_{k-L}, &  k > L.  \end{cases}
\end{align*}
Note that the rates are identical to those achieved using the construction on $D_4^*$.

We now show that this lattice index code provides uniform side information gain of \mbox{$\Gamma \approx 6$~dB/b/dim}. The proof is similar to the proofs of Lemmas~\ref{lem:arbit_construction} and~\ref{lem:dS_RS} in Section~\ref{sec:commutative}.

\begin{lemma} \label{lem:quat_lattices_injective}
With $\La_1,\dots,\La_K$ and $\Ls$ defined as above,
\begin{enumerate}[\itshape(i\itshape)]
\item the transmit codebook $\mathscr{C}=\La/\Ls$, and the encoding map $\rho$ is injective; and
\item for every side information index set $S$, $R_S=\log_2\left( \sfrac{d_S}{d_0} \right)$.
\end{enumerate}
\end{lemma}
\begin{IEEEproof}
See Appendix~\ref{app:lem:quat_lattices_injective}.
\end{IEEEproof}

From Lemma~\ref{lem:quat_lattices_injective}, we conclude that the side information gain of the quaternionic lattice index code $\La/\Ls$ is \mbox{$\sim 6$~dB/b/dim}.

\section{Coding for General Message Demands: An Example} \label{sec:3receiver}

Lattice index codes with large side information gains are suitable when all the messages are demanded by every receiver. For these codes, the encoding operation is oblivious to both the number of receivers and the side information configuration at each receiver (see Definition~\ref{def:lattice_index_code}). When the message demands are more general (such as private message requests), the number of receivers, and the ${\sf SNR}$ and the side information available at each receiver may need to be considered during code design~\cite{SiC_ISIT_14,AOJ_ISIT_14}.

Capacity-achieving random coding schemes have been proposed for a class of $3$-receiver private message Gaussian broadcast channels in~\cite{SiC_ISIT_14} and~\cite{AOJ_ISIT_14}. The coding schemes of~\cite{AOJ_ISIT_14} make use of channel codes that are efficient in converting receiver side information into additional coding gains, similar to lattice index codes, as component subcodes in superposition coding.
In this section, we consider an instance of a broadcast channel where each message is demanded at a unique receiver. Inspired by the ideas in~\cite{AOJ_ISIT_14}, we show that lattice index codes with large side information gains can be useful in constructing coding schemes that are matched to this broadcast channel.

We will now briefly review some lattice parameters from~\cite{CoS_Springer_99} that are relevant to the analysis of error performance.
The \emph{kissing number} $\tau(\La)$ of a lattice $\La$ is the number of shortest non-zero vectors in $\La$, i.e., the number of lattice points with Euclidean length equal to $d_{\min}(\La)$. Every point in $\La$ has exactly $\tau(\La)$ nearest neighbours in the lattice.
The \emph{covering radius} of a lattice $\La$ is given by
\begin{equation} \label{eq:cov_radius}
 r_{\rm cov}\left(\La\right) = \sup_{x \in \mathcal{V}_{\La}} \|x\|,
\end{equation}
where $\mathcal{V}_{\La}$ is the fundamental Voronoi region of $\La$, and equals the radius of the smallest sphere centered around origin that contains the fundamental Voronoi region as a subset.

\begin{figure*}[!t]
\centering
\includegraphics[width=5in]{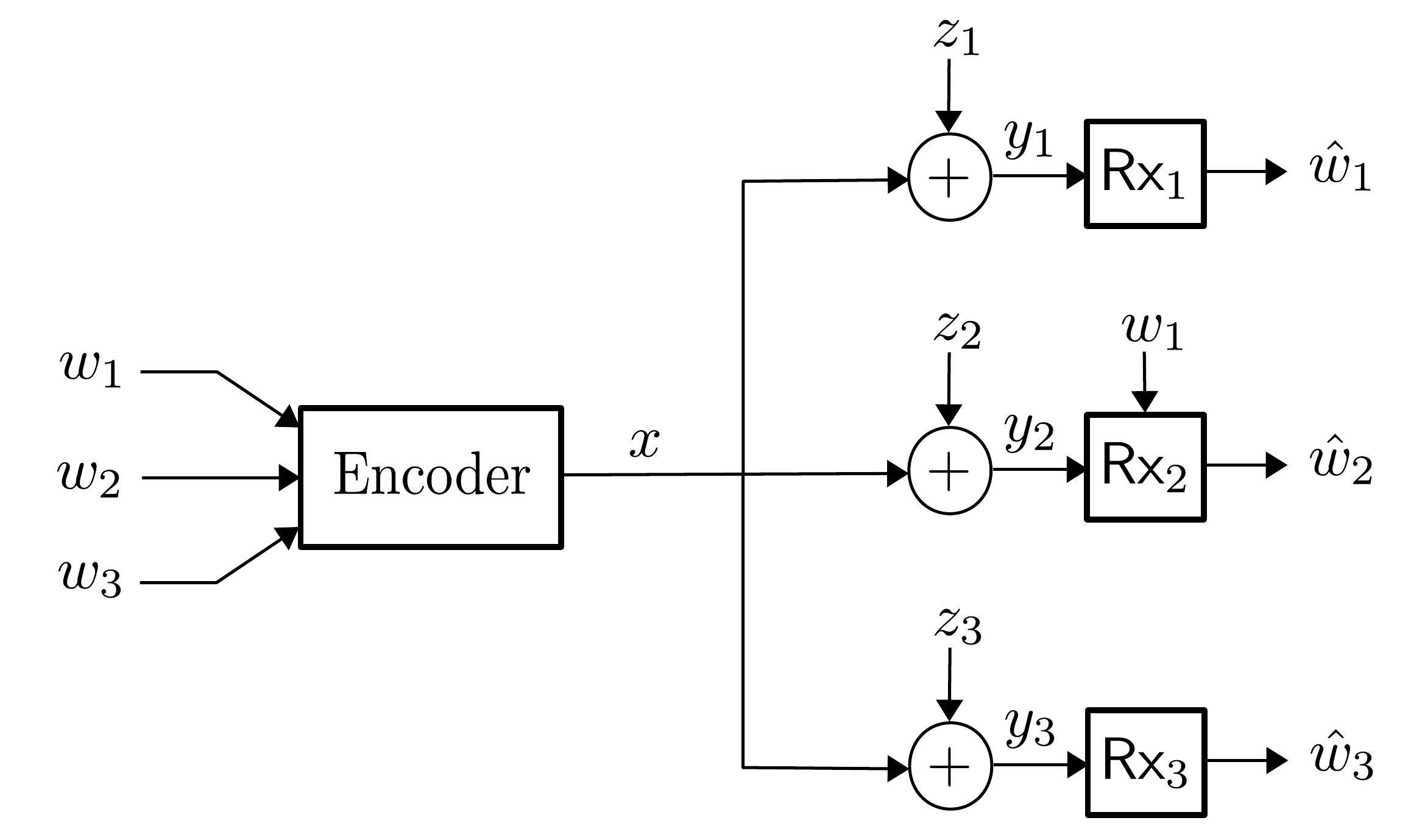}
\caption{A three receiver Gaussian broadcast channel with private message requests and side information at ${\sf Rx}_2$.}
\label{fig:3_receiver}
\end{figure*}

\subsection{Channel model and Encoding}

We consider a broadcast channel with three receivers ${\sf Rx}_j$, \mbox{$j=1,2,3$}, each of which experiences additive noise with the corresponding variance $N_j$, see Fig.~\ref{fig:3_receiver}. We assume that \mbox{$N_1 \leq N_2 \leq N_3$}, i.e., the first receiver has the strongest channel.
Also assume that there are \mbox{$K=3$} messages at the transmitter, \mbox{$w_k \in \mathcal{W}_k$}, \mbox{$k=1,2,3$}. Let \mbox{$\Dc_j,S_j \subset \{1,2,3\}$} denote the index sets of the messages demanded by, and the side information available at ${\sf Rx}_j$. We consider the private message broadcast scenario $\Dc_1=\{1\}$, $\Dc_2=\{2\}$, $\Dc_3=\{3\}$, with side information index sets $S_1=\varnothing$, $S_2=\{1\}$, $S_3=\varnothing$.

The objective is to efficiently encode the messages such that the three receivers ${\sf Rx}_1,{\sf Rx}_2,{\sf Rx}_3$ can tolerate increasingly more noise, i.e., the messages $w_1,w_2,w_3$ experience increasing coding gains, in that order.
Using a lattice index code, we will exploit the side information $S_2$ to enhance the coding gain of ${\sf Rx}_2$ over that of ${\sf Rx}_1$.
Since $S_3=\varnothing$, we will combine this lattice index code with superposition coding to enhance the coding gain at ${\sf Rx}_3$.

The transmitter uses nested lattices \mbox{$\La_1,\La_2 \supset \Lscomb$} and \mbox{$\La_3 \supset \Lsth$}, to individually map the information symbols $w_1,w_2,w_3$ to the points $x_1,x_2,x_3$ in the $n$-dimensional lattice constellations $\La_1/\Lscomb$, $\La_2/\Lscomb$ and $\La_3/\Lsth$, respectively. Finally, the transmit vector is generated as
\begin{equation*}
 x = (x_1 + x_2) \modc \Lscomb + x_3 = x_{12} + x_3,
\end{equation*}
where $x_{12}=(x_1 + x_2) \modc \Lscomb$.
We assume that the map $(x_1,x_2) \to (x_1+x_2) \modc \Lscomb$ generates a lattice index code $\Cc_{12}=\Lacomb/\Lscomb$, where \mbox{$\Lacomb=\La_1+\La_2$} denotes the sum lattice.
Denoting $\Lath/\Lsth$ by $\Cc_3$, we observe that the transmit codebook $\Cc=\Cc_{12}+\Cc_3$ is a superposition code, where the codewords of $\Cc_{12}$ form the `cloud particles' and those of $\Cc_3$ are the `cloud centers'~\cite{CoT_JohnWiley_12}.

\subsection{Decoding and Error Performance}

The weakest receiver ${\sf Rx}_3$ observes $y_3 = x_{12} + x_3 + z_3$, where $z_3$ is a random Gaussian vector with variance $N_3$ per dimension. The optimal decoder chooses $\hat{x}_3 \in \Lath/\Lsth$ that maximizes the likelihood of observing $y_3$. Since this receiver is complex to analyze, we consider the sub-optimal decoder that treats the `interference' $x_{12}$ as noise, and decodes $y_3$ to the nearest point in $\Lath/\Lsth$. We now derive an upper bound on the pairwise error probability of this receiver considering two competing codewords $x_A,x_B \in \Lath/\Lsth$.
Assuming that $w_3$ was encoded as \mbox{$x_A \in \Lath$}, the decoder at ${\sf Rx}_3$ chooses $x_B \in \Lath$ over $x_A$ if $\|y-x_A\| > \|y-x_B\|$, i.e., if
\begin{equation*}
\| x_{12} + x_A + z_3 - x_A \| > \|x_{12} + x_A + z_3 - x_B\|,
\end{equation*}
where \mbox{$x_{12} \in \Cc_{12}$} is the vector that jointly encodes $w_1,w_2$. Squaring both sides of the inequality and using usual simplifications, we arrive at
\begin{equation*}
2 z_3^\intercal(x_B-x_A) > \|x_A-x_B+x_{12}\|^2 - \|x_{12}\|^2.
\end{equation*}
To upper bound the error probability, we obtain a lower bound on the value of the right-hand-side term above. Utilizing the Cauchy-Schwarz inequality, we obtain
\begin{align*}
|x_A&-x_B+x_{12}\|^2 - \|x_{12}\|^2\\
&= \|x_A-x_B\|^2 + \|x_{12}\|^2 + 2x_{12}^\intercal(x_A-x_B) - \|x_{12}\|^2 \\
&= \|x_A-x_B\|^2 + 2x_{12}^\intercal(x_A-x_B) \\
&\geq \|x_A - x_B\|^2 - 2 \left\vert x_{12}^\intercal(x_A-x_B) \right\vert \\
&\geq \|x_A-x_B\|^2 - 2 \|x_{12}\| \, \|x_A-x_B\| \\ &= \|x_A-x_B\| \left( \|x_A-x_B\| - 2\|x_{12}\|\right).
\end{align*}
Observe that \mbox{$x_{12} \in \Lacomb/\Lscomb$}, and hence, \mbox{$x_{12} \in \mathcal{V}_{\Lscomb}$}. From the definition of the covering radius~\eqref{eq:cov_radius}, we have \mbox{$\|x_{12}\| \leq r_{\rm cov}(\Lscomb)$}. Since \mbox{$x_A,x_B \in \Lath$}, we have \mbox{$\|x_A-x_B\| \geq d_{\min}(\Lath)$}. This yields the following lower bound
\begin{align*}
 \|x_A-x_B+x_{12}&\|^2 - \|x_{12}\|^2 \\ \geq &\|x_A-x_B\| \, \left( d_{\min}(\Lath) - 2 r_{\rm cov}\left(\Lscomb\right) \right).
\end{align*}
Hence, ${\sf Rx}_3$ favours $x_B$ \emph{only if} $z_3$ is such that
\begin{equation*}
 2 z_3^\intercal(x_B-x_A) > \|x_A-x_B\| \, \left( d_{\min}(\Lath) - 2 r_{\rm cov}\left(\Lscomb\right) \right) .
\end{equation*}
Normalizing both sides by $2\sqrt{N_3}\,\|x_A-x_B\|$, we immediately obtain the following upper bound on pairwise error probability,
\begin{align*}
 {\sf PEP}(x_A \to x_B) \leq Q\left( \frac{d_{\min}(\Lath) - 2 r_{\rm cov}\left(\Lscomb\right)}{2\sqrt{N_3}} \right),
\end{align*}
where $Q(\cdot)$ is the Gaussian tail function and $N_3$ is the variance of the vector $z_3$ along each dimension.

An approximate bound on the average error probability can be obtained by considering all the competing codewords which are at the shortest Euclidean distance from the transmitted codeword~\cite{CoS_Springer_99}, i.e., all the nearest neighbours in the coding lattice. Using union bound, we arrive at the following approximate bound~\cite{CoS_Springer_99} for error rate at ${\sf Rx}_3$
\begin{align} \label{eq:pe3}
P_e({\sf Rx}_3) &\lesssim \tau\left( \Lath \right) \, {\sf PEP}(x_A \to x_B) \nonumber \\ &\leq \tau\left( \Lath \right) \, Q\left( \frac{d_{\min}(\Lath) - 2 r_{\rm cov}\left(\Lscomb\right)}{2\sqrt{N_3}} \right).
\end{align}

To analyze the performance at ${\sf Rx}_1$ and ${\sf Rx}_2$, we again consider sub-optimal decoders for which upper bounds on error probabilities can be easily obtained. The decoders at ${\sf Rx}_1$ and ${\sf Rx}_2$ experience a higher ${\sf SNR}$ than ${\sf Rx}_3$. Both these receivers first decode $w_3$ using the same procedure as ${\sf Rx}_3$, and subtract its contribution in the received vector. Assuming that the estimated codeword $\hat{x}_3$ is correct, the received vector at ${\sf Rx}_j$, $j=1,2$, after cancelling the interference $x_3$ is
\begin{align*}
y_j'=x_{12} + z_j = (x_1 + x_2) \modc \Lscomb + z_j,
\end{align*}
where $z_j$ is a Gaussian noise vector with variance $N_j$ per dimension. Since ${\sf Rx}_1$ has no side information, it jointly decodes $w_1$ and $w_2$, i.e., it chooses the codeword $\hat{x}_{12} \in \Lacomb/\Lscomb$ that is closest to $y_1'$.
Using conventional union bounding arguments, the overall error probability at this receiver, considering both the steps of the decoding procedure, can be upper bounded as
\begin{align}
P_e({\sf Rx}_1) \lesssim \tau & \left( \Lacomb \right) \, Q\left( \frac{d_{\min}(\Lacomb)}{2\sqrt{N_1}} \right) \nonumber \\ &~~+ \, \tau\left( \Lath \right) \, Q\left( \frac{d_{\min}(\Lath) - 2 r_{\rm cov}\left(\Lscomb\right)}{2\sqrt{N_1}} \right).\label{eq:pe1}
\end{align}
On the other hand, ${\sf Rx}_2$ has prior knowledge of the exact value $a_1$ of $x_1$ and its decoder can exploit the fact that $\Lacomb/\Lscomb$ is a lattice index code. The effective codebook seen by this receiver after cancelling the interference $x_3$ and expurgating all codewords corresponding to $x_1 \neq a_1$ is a lattice code carved from a translate of $\La_2$. Hence, the error rate at this receiver satisfies
\begin{align}
P_e({\sf Rx}_2) \lesssim \tau & \left( \La_2 \right) \, Q\left( \frac{d_{\min}(\La_2)}{2\sqrt{N_2}} \right) \nonumber \\ &~~+ \, \tau\left( \Lath \right) \, Q\left( \frac{d_{\min}(\Lath) - 2 r_{\rm cov}\left(\Lscomb\right)}{2\sqrt{N_2}} \right).\label{eq:pe2}
\end{align}

At high values of ${\sf SNR}$, the arguments of the $Q$-function in~\eqref{eq:pe3},~\eqref{eq:pe1} and~\eqref{eq:pe2} dictate the error performance at the three receivers. Since ${\sf Rx}_3$ experiences the most noise, we require $d_{\min}(\Lath)-2\,r_{\rm cov}(\Lscomb)$ to be larger than $d_{\min}(\La_2)$ and $d_{\min}(\Lacomb)$. In this case, the high ${\sf SNR}$ error rates at the three receivers ${\sf Rx}_1,{\sf Rx}_2,{\sf Rx}_3$ are determined by $d_{\min}(\Lacomb)$, $d_{\min}(\La_2)$ and $d_{\min}(\Lath)-2r_{\rm cov}(\Lscomb)$, respectively. Hence, we arrive at the following guidelines for designing a good channel code:
\begin{enumerate}[\itshape(i\itshape)]
\item $\Lacomb/\Lscomb$ must be a good lattice index code in order to achieve a good error performance at ${\sf Rx}_1$ and ${\sf Rx}_2$.
A large value of $\Gamma(\Lacomb/\Lscomb)$ will be efficient in converting the side information into additional coding gains, which will be useful in combating the higher noise power at ${\sf Rx}_2$.
\item The covering radius of $\Lscomb$ must be small, so as to reduce the interference from $x_{12}$ at ${\sf Rx}_3$.
\item And finally, $d_{\min}(\Lath)$ must be large in order to maximize the coding gain at ${\sf Rx}_3$.
\end{enumerate}

\begin{example}
We will consider a coding scheme for the $3$-user private message broadcast channel that utilizes the $25$-QAM constellation of Example~\ref{ex:25QAM} as the lattice index code $\Lacomb/\Lscomb$. This constellation has dimension \mbox{$n=2$} and encodes two messages with $5$-ary alphabets. From Example~\ref{ex:25QAM}, we have \mbox{$d_{\min}(\La_{12})=1$} and \mbox{$d_{\min}(\La_2)=\sqrt{5}$}. To encode the third message, we will use \mbox{$\Lsth=25\Zb^2$}, and the lattice generated by
\begin{equation*}
\begin{pmatrix} 10 & -5 \\ 5 & 10 \end{pmatrix}
\end{equation*}
as $\Lath$. It is straightforward to show that \mbox{$r_{\rm cov}(\Lscomb)=\sfrac{5}{\sqrt{2}}$}, \mbox{$d_{\min}(\Lath)=5\sqrt{5}$}, and that all three messages are encoded at the same rate \mbox{$R_1=R_2=R_3=\sfrac{1}{2} \log_2 5$~b/dim}. At high ${\sf SNR}$, the error performance at ${\sf Rx}_2$ is better than ${\sf Rx}_1$ by
\begin{equation*}
10 \log_{10} \left( \frac{d_{\min}^2(\La_2)}{d_{\min}^2(\La_{12})} \right) = 6.9 \text{ dB},
\end{equation*}
and the performance at ${\sf Rx}_3$ is better than ${\sf Rx}_1$ by
\begin{equation*}
10 \log_{10} \left( \frac{\left( d_{\min}(\Lath) - 2\,r_{\rm cov}(\Lscomb) \right)^2}{d_{\min}^2(\Lacomb)} \right) = 12.2 \text{ dB}.
\end{equation*}
Hence, this constellation allows ${\sf Rx}_2$ and ${\sf Rx}_3$ to tolerate $6.9$~dB and $12.2$~dB of additional noise compared to ${\sf Rx}_1$, respectively.
While the additional gain at ${\sf Rx}_3$ is due to superposition coding, the performance improvement at ${\sf Rx}_2$ is due to the side information gain of the component lattice index code.
\hfill\IEEEQED
\end{example}

\section{Conclusion and Discussion}

We have proposed lattice index codes for the Gaussian broadcast channel where every receiver demands all the messages from the transmitter. We have introduced the notion of side information gain as a code design metric, and constructed lattice index codes from lattices $\La$ over the PIDs $\Zb$, $\Zb[i]$, $\Zb[\omega]$ and $\Hb$. If $\La$ has the highest lattice density in its dimension, the proposed codes achieve the maximum side information gain among all lattice index codes constructed from $\La$. An interesting property of these lattice index codes is that the side information gain is uniform.

The key ingredients that we used in the construction of our lattice index  codes are the Chinese remainder theorem,
the properties of principal ideals for the base PIDs, and the mapping of ideals of the PID modules to lattice constellations.
In particular, the specific choices of the PIDs enable us to associate the norms of principal ideals with the minimum Euclidean distance of the corresponding component lattices, while the Chinese remainder theorem guarantees the unique decodability property for any amount of side information at the receivers.

It is possible to construct lattice index codes using the $8$-dimensional non-commutative non-associative PID of Octavian integers $\mathbb{O}$. Since $\mathbb{O}$ is geometrically equivalent to the Gosset lattice $E_8$, the resulting lattice index codes use the octonion version of $E_8$ as the base lattice $\Lt$. However, the only ideals in $\mathbb{O}$ are the trivial ones, viz. the ideals $m\mathbb{O}$, where \mbox{$m \in \Zb$}~\cite{CoS_Peters_03}. Hence the extension of our construction from the Hurwitz integers $\Hb$ to the Octavian integers $\mathbb{O}$ coincides with the codes constructed in Section~\ref{sec:commutative} with $\La=E_8$ and $\Db=\Zb$.

The lattice index codes constructed here can be used as modulation schemes together with strong outer codes. Consider $K$ information streams, encoded independently using $K$ outer codes over the alphabets $\mathcal{W}_1,\dots,\mathcal{W}_K$, respectively. The coded information streams are multiplexed using the lattice index code $\mathscr{C}$ and transmitted. If the minimum Hamming distance of the outer codes is $d_{H}$, then the minimum squared Euclidean distance at a receiver corresponding to $S$ is at least \mbox{$d_H \times d_S^2$}. While the outer code improves error resilience, the inner lattice index code collects the gains from side information.
This approach converts the index coding problem into coding for a multiple-access channel where the $K$ information streams are viewed as $K$ independent transmitters. Since coding for multiple-access channels is well studied in the literature, this knowledge may be leveraged to construct good noisy index codes of manageable encoding and decoding complexity, such as by using iterative multiuser demodulators/decoders. In~\cite{NHV_ISIT_15} we have shown that this concatenated architecture can perform close to the capacity of the Gaussian broadcast channel with receiver side information.


\appendices

\section{Proofs of Lemmas}

\subsection{Proof of Lemma~\ref{lem:arbit_construction}} \label{app:lem:arbit_construction}
In order to prove Part~\emph{(\ref{lem:part1:arbit_construction})}, we need to show that $\rho$ is injective and $\La_1+\cdots+\La_K=\La$.

From Lemma~\ref{lem:gcd_M_S}, \mbox{$\gcd(M_k,k \in S^\mathsf{c})=\prod_{\ell \in S}\phi_{\ell}$} for every choice of $S$. Hence, there exists a tuple \mbox{$(b_k,k \in S^\mathsf{c})$} of elements in $\Db$ such that
$\sum_{k \in S^\mathsf{c}}b_kM_k=\prod_{\ell \in S}\phi_{\ell}$.
It follows that, for every \mbox{$\lambda \in \Lt$}, we have
$$\prod_{\ell \in S}\phi_{\ell} \lambda=\sum_{k \in S^\mathsf{c}}b_kM_k\lambda,$$
hence
\mbox{$\prod_{\ell \in S}\phi_{\ell} \, \Lt \subset \sum_{k \in S^{\mathsf{c}}}M_k \Lt$}. Using this result along with the additive property of $\Psi$, we obtain
\begin{align*}
\Psi\left( \prod_{\ell \in S}\phi_{\ell} \, \Lt \right) \subset \Psi\left(\sum_{k \in S^{\mathsf{c}}}M_k \Lt\right) &= \sum_{k \in S^{\mathsf{c}}} \Psi\left( M_k \Lt \right) \\ &= \sum_{k \in S^{\mathsf{c}}} \La_k.
\end{align*}
Considering cosets modulo $\Ls$, the above relation implies
\begin{align} \label{eq:arbit_lattice_1}
\Psi\left( \prod_{\ell \in S}\phi_{\ell} \, \Lt \right) \,/\Ls \subset \sum_{k \in S^{\mathsf{c}}} \La_k/\Ls.
\end{align}

Let $\rho\vert_{S^{\mathsf{c}}}$ be the restriction of the encoding map~\eqref{eq:rho} to the message symbols with indices in $S^{\mathsf{c}}$, i.e.,
\begin{align*}
\rho\vert_{S^{\mathsf{c}}} \left( x_k, k \in S^{\mathsf{c}} \right) = \sum_{k \in S^{\mathsf{c}}} x_k {\rm~mod~} \Ls.
\end{align*}
Note that $\sum_{k \in S^\mathsf{c}}\La_k/\Ls$ is the image of the map $\rho\vert_{S^\mathsf{c}}$. From~\eqref{eq:arbit_lattice_1}, we observe that $\Psi\left( \prod_{\ell \in S}\phi_{\ell} \Lt \right)/\Ls$ is a subset of this image. The cardinality
\begin{align*}
\left|\Psi\left( \prod_{\ell \in S}\phi_{\ell} \Lt \right)/\Ls\right| = \frac{ |M|^n \vol(\La) }{ |\prod_{\ell \in S} \phi_{\ell}|^n \vol(\La)} = \prod_{k \in S^{\mathsf{c}}} \left| \phi_k \right|^n
\end{align*}
of this subset of the image of $\rho\vert_{S^{\mathsf{c}}}$ equals the cardinality
\begin{align*}
\prod_{k \in S^\mathsf{c}} \left|\La_k/\Ls\right| = \prod_{k \in S^{\mathsf{c}}} \left| \phi_k \right|^n
\end{align*}
of the domain of $\rho\vert_{S^{\mathsf{c}}}$. Hence, we conclude that $\rho\vert_{S^{\mathsf{c}}}$ is an injective map, and the subset $\Psi\left( \prod_{\ell \in S}\phi_{\ell} \Lt \right)/\Ls$ equals the entire image $\sum_{k \in S^{\mathsf{c}}} \La_k/\Ls$. This implies that $\Psi\left( \prod_{\ell \in S}\phi_{\ell} \Lt \right)=\sum_{k \in S^{\mathsf{c}}} \La_k$, proving Part~\emph{(\ref{lem:part3:arbit_construction})} of this lemma.

Choosing \mbox{$S=\varnothing$}, we observe that \mbox{$\rho\vert_{S^{\mathsf{c}}}=\rho$} is injective, and \mbox{$\sum_{k=1}^{K}\La_k = \Psi\left( \Lt \right)=\La$}. Hence, the transmit codebook is
\mbox{$\mathscr{C}=\sum_{k=1}^{K} \La_k/\Ls = \La/\Ls$}. 
This proves Part~\emph{(\ref{lem:part1:arbit_construction})}.
\hfill\IEEEQED

\subsection{Proof of Lemma~\ref{lem:H1}} \label{app:lem:H1}

It is enough to show that \mbox{$\La=\Hb$}, i.e., \mbox{$\sum_{k=1}^{K} \Hb M_k = \Hb$}, or equivalently, $$\gcd(M_1,\dots,M_K)=1.$$ Let \mbox{$D=\gcd(M_1,\dots,M_K)$} and $D_k=\gcd(M_k,M_{k+L})$ for \mbox{$k=1,\dots,L$}. Then,
\begin{align}
 D &= \gcd(M_1,M_{1+L},M_2,M_{2+L},\dots,M_L,M_{2L}) \nonumber \\
   &= \gcd\left( \gcd(M_1,M_{1+L}), \dots,\gcd(M_L,M_{2L}) \right) \nonumber \\
   &= \gcd\left( D_1,\dots,D_L \right). \label{eq:gcd_H_12}
\end{align}
We will complete the proof by deriving $N(D_1),\dots,N(D_L)$, and then showing that $D$ is a unit in $\Hb$.

For each \mbox{$k=1,\dots,L$}, we have
\begin{align*}
       D_k &= \gcd(M_k,M_{k+L}) = \gcd(M_k,M_k+M_{k+L}) \\ 
                  &= \gcd\left( P_k \prod_{\ell \neq k}p_{\ell}, P_k \prod_{\ell \neq k}p_{\ell} + \overline{P}_k \prod_{\ell \neq k} p_{\ell}  \right) \\
                  &= \gcd\left( P_k \prod_{\ell \neq k}p_{\ell}, 2^{m+1} \prod_{\ell \neq k} p_{\ell}  \right),
\end{align*}
where the last equality follows from the assumption that {$\real(P_k)=2^m$} for some {$m \geq 0$}. Since
\begin{align*}
N(D_k)\divides \gcd(N(M_k),N(M_{k}+M_{k+L})),
\end{align*}
we obtain {$N(D_k) \divides  \gcd\left( p_k \prod_{\ell \neq k} p_{\ell}^2, 4^{m+1} \prod_{\ell \neq k} p_{\ell}^2  \right)$}. Since $p_k$ is an odd prime, we have
\begin{align} \label{eq:gcd_H_1}
N(D_k) \divides  \prod_{\ell \neq k}p_{\ell}^2.
\end{align}
On the other hand, $\prod_{\ell \neq k}p_{\ell}$ is a divisor of both $M_k$ and $M_{k+L}$, and hence is a divisor of $D_k$. Hence, 
\begin{align} \label{eq:gcd_H_2}
N\left(\prod_{\ell \neq k}p_{\ell}\right)\divides N(D_k), \text{ i.e., } \prod_{\ell \neq k}p_{\ell}^2\divides N(D_k).
\end{align}
From~\eqref{eq:gcd_H_1} and~\eqref{eq:gcd_H_2}, \mbox{$N(D_k)=\prod_{\ell \neq k}p_{\ell}^2$}. 

From~\eqref{eq:gcd_H_12}, $N(D)\divides \gcd(N(D_1),\dots,N(D_L))$ in $\Zb$. Since $p_1,\dots,p_L$ are pairwise relatively prime in $\Zb$,
\begin{align*}
\gcd(N(D_1),\dots,N(D_L))
=\gcd\left( \prod_{\ell \neq 1}p_{\ell}^2,\dots,\prod_{\ell \neq L}p_{\ell}^2  \right)
=1.
\end{align*}
Hence \mbox{$N(D)=1$}, and $D$ is a unit in $\Hb$. Up to unit multiplication in $\Hb$, we have
\begin{align} \label{eq:gcd_Mk_is_1}
D = \gcd(M_1,\dots,M_K) = 1.
\end{align}
\hfill\IEEEQED

\subsection{Proof of Lemma~\ref{lem:quat_lattices_injective}} \label{app:lem:quat_lattices_injective}

\subsubsection*{Part~(i)}

It is enough to show that \mbox{$\sum_{k=1}^{K}\La_k=\La$}, or equivalently, $\sum_{k=1}^{K}\Lt_k=\Lt$.
Since $\Lt_k \subset \Lt$, for all $k$, it is clear that
$$\sum_{k=1}^{K} \Lt_k \subset \Lt.$$
From~\eqref{eq:gcd_Mk_is_1}, we have $\gcd(M_1,\dots,M_K)=1$. Hence, there exist $B_1,\dots,B_K \in \Hb$ such that $\sum_{k=1}^{K}B_kM_k=1$. If $\lambda \in \Lt$, then
\begin{align*}
\lambda = \lambda\sum_{k=1}^{K}B_kM_k = \sum_{k=1}^{K} \left( \lambda B_k \right) M_k.
\end{align*}
Since $(\lambda B_k)M_k \in \Lt_k$, we have $\lambda \in \sum_{k=1}^{K}\Lt_k$. Hence
$$\Lt \subset \sum_{k=1}^{K} \Lt_k.$$
The injective nature of the map $\rho$ follows from observing that its domain $\La_1/\Ls \times \cdots \times \La_K/\Ls$ and image $\La/\Ls=\Psi(\Lt)/\Psi(\Lt M)$ have the same cardinality $N(M)^{2t}=\left( \prod_{\ell=1}^{L}p_{\ell}^{2t} \right)^2$.


\subsubsection*{Part~(ii)}

Let $D_S=\gcd(M_k,k \in S^{\mathsf{c}})$. We first show that $\sum_{k \in S^{\mathsf{c}}}\La_k=\Psi(\Lt D_S)$, or equivalently $\sum_{k \in S^{\mathsf{c}}}\Lt_k = \Lt D_S$. There exists a tuple $(B_k,k \in S^{\mathsf{c}})$ of Hurwitz integers such that $\sum_{k \in S^{\mathsf{c}}}{B_kM_k}=D_S$. Similar to the proof of Part~\emph{(i)} of this lemma, by considering the term $\lambda\sum_{k \in S^{\mathsf{c}}}B_kM_k$ for each $\lambda \in \Lt$, we conclude that
\begin{align*}
\sum_{k \in S^{\mathsf{c}}} \Lt_k \supset \Lt D_S.
\end{align*}
The above relation implies that $\Psi(\Lt D_S)/\Ls$ is a subset of the image of $\rho\vert_{S^{\mathsf{c}}}$, which is the restriction of the function $\rho$ to messages with indices in $S^{\mathsf{c}}$. As in the proof of Lemma~\ref{lem:arbit_construction}, to prove $\sum_{k \in S^{\mathsf{c}}} \Lt_k = \Lt D_S$, it is enough to show that
\begin{align*}
|\Psi(\Lt D_S)/\Ls| = \prod_{k \in S^{\mathsf{c}}} |\La_k/\Ls|.
\end{align*}
Now,
\begin{align*}
N(M)^{2t} = \frac{\vol(\Lt M)}{\vol(\Lt)} &= |\Psi(\Lt)/\Psi(\Lt M)| \\  &= |\La/\Ls| = |\mathscr{C}| = 2^{4t\left( R_1 + \cdots + R_K \right)}.
\end{align*}
Using $N(D_S)^2=2^{4R_S}$ (from~\eqref{eq:norm_DS}), and the above equation, we have
\begin{align*}
|\Psi(\Lt D_S)/\Ls| &= \frac{\vol(\Lt M)}{\vol(\Lt D_S)} = \frac{N(M)^{2t}}{N(D_S)^{2t}} = \frac{2^{4t(R_1+\cdots+R_K)}}{2^{4tR_S}} \\
 &= 2^{4t \sum_{k \in S^{\mathsf{c}}}R_k} = \prod_{k \in S^{\mathsf{c}}} 2^{4tR_k} =  \prod_{k \in S^{\mathsf{c}}}|\La_k/\Ls|.
\end{align*}
Hence, we conclude that $\sum_{k \in S^{\mathsf{c}}} \Lt_k = \Lt D_S$.

Using $N(D_S)=2^{2R_S}$, we obtain the minimum squared distance with $S$ as follows,
\begin{align*}
d_S^2 &= d_{\min}^2\left(\sum_{k \in S^{\mathsf{c}}} \La_k \right) = d_{\min}^2\left(\sum_{k \in S^{\mathsf{c}}}\Lt_k\right) \\ &= d_{\min}^2\left( \Lt D_S \right) = N(D_S) \, d_{\min}^2(\Lt) = 2^{2R_S} d_0^2.
\end{align*}
This shows that $R_S=\log_2\left( \sfrac{d_S}{d_0}\right)$.
\hfill\IEEEQED

\section{Existence of Hurwitz integers with odd-prime norms and real part a power of two} \label{app:hurwitz_prime_real_part}

We show that every odd rational prime $p$ can be expressed as the sum of the squares of four rational integers $a_1,\dots,a_4$, where the first integer $a_1 \in \{1,2\}$. Then, $P=a_1 + a_2 i + a_3 j + a_4 k$ is a Hurwitz integer with norm $p$ and real part a power of $2$.
The proof follows from the following result from number theory known as the three-square theorem.

\begin{theorem}[\cite{UsH_McGraw_39}]
Every positive rational integer not of the form $4^c(8d+7)$, $c,d \in \Zb$, is a sum of three rational integer squares.
\end{theorem}

If $p$ is a positive odd rational integer, we have $p {\rm~mod~} 8 \in \{1,3,5,7\}$. For each of these four possible values of $p {\rm~mod~} 8$, we show that at least one of $p-1$ or $p-4$ is not of the form $4^c (8d+7)$. It then follows that, either $p-1$ or $p-4$ is a sum of three squares, and consequently, $p$ equals either the sum of $1^2$ and three squares, or the sum of $2^2$ and three squares.

If \mbox{$p {\rm~mod~8}=1$}, then
$$(p-4) {\rm~mod~} 8= \left( p {\rm~mod~}8 - 4 \right) {\rm~mod~}8=5.$$
Assume $p-4=4^c(8d+7)$ for some $c,d \in \Zb$. Since \mbox{$(p-4) {\rm~mod~}8=5$}, $(p-4)$ is odd, which implies $c=0$, and hence, $p-4=8d+7$. This leads to a contradiction since $(p-4) {\rm~mod~} 8=5$ and $(8d+7){\rm~mod~} 8=7$.
The proofs for the cases $p {\rm~mod~}8=5,7$ are similar.

If $p {\rm~mod~}8=3$, we have $(p-1){\rm~mod~}8=2$. Suppose $p-1=4^c(8d+7)$ for some choice of $c,d$. Since $(p-1){\rm~mod~}8 \notin \{0,4\}$, $4$ is not a divisor of $p-1$, and hence, $c=0$. Contradiction follows from observing that $(p-1) {\rm~mod~}8 \neq (8d+7) {\rm~mod~}8$.



\end{document}